\documentclass[twocolumn,showpacs,superscriptaddress,amsmath,amssymb,aps,prl]{revtex4-1}

\usepackage{mathptmx}

\usepackage[latin9]{inputenc}
\usepackage{amsmath}
\usepackage{amssymb}
\usepackage{graphicx}
\usepackage{esint}
\usepackage{times}
\usepackage{array}
\usepackage{ulem}

\usepackage[unicode=true,
 bookmarks=true,bookmarksnumbered=false,bookmarksopen=false,
 breaklinks=false,pdfborder={0 0 1},backref=false,colorlinks=true]
 {hyperref}
\hypersetup{linkcolor=magenta, urlcolor=blue, citecolor=blue, pdfstartview={FitH}, hyperfootnotes=false, unicode=true}

\makeatletter

\pdfpageheight\paperheight
\pdfpagewidth\paperwidth

\@ifundefined{textcolor}{}
{%
 \definecolor{BLACK}{gray}{0}
 \definecolor{WHITE}{gray}{1}
 \definecolor{RED}{rgb}{1,0,0}
 \definecolor{GREEN}{rgb}{0,1,0}
 \definecolor{BLUE}{rgb}{0,0,1}
 \definecolor{CYAN}{cmyk}{1,0,0,0}
 \definecolor{MAGENTA}{cmyk}{0,1,0,0}
 \definecolor{YELLOW}{cmyk}{0,0,1,0}
}

\usepackage{xcolor}\usepackage{soul}

\definecolor{blue}{rgb}{0,0,1}
\definecolor{red}{rgb}{1,0,0}
\definecolor{green}{rgb}{0,1,0}

\makeatother

\begin{document}
\bibliographystyle{unsrt}
\normalem

\title{Exceptional Point and Cross-Relaxation Effect in a Hybrid Quantum System}
\author{Guo-Qiang Zhang}
\thanks{These authors contributed equally to this work.}
\affiliation{Interdisciplinary Center of Quantum Information, State Key Laboratory of Modern Optical Instrumentation, and Zhejiang Province Key Laboratory of Quantum Technology and Device, Department of Physics, Zhejiang University, Hangzhou 310027, China}

\author{Zhen Chen}
\thanks{These authors contributed equally to this work.}
\affiliation{Interdisciplinary Center of Quantum Information, State Key Laboratory of Modern Optical Instrumentation, and Zhejiang Province Key Laboratory of Quantum Technology and Device, Department of Physics, Zhejiang University, Hangzhou 310027, China}
\affiliation{Beijing Academy of Quantum Information Sciences, Beijing 100193, China}

\author{Da Xu}
\affiliation{Interdisciplinary Center of Quantum Information, State Key Laboratory of Modern Optical Instrumentation, and Zhejiang Province Key Laboratory of Quantum Technology and Device, Department of Physics, Zhejiang University, Hangzhou 310027, China}

\author{Nathan Shammah}
\affiliation{Theoretical Quantum Physics Laboratory, RIKEN Cluster for Pioneering Research, Wako-shi, Saitama 351-0198, Japan}
\affiliation{Unitary Fund, Walnut, CA 91789, USA}
\affiliation{Quantum Technology Lab, Dipartimento di Fisica, Universit\`{a} degli Studi di Milano, 20133 Milano, Italy}

\author{Meiyong Liao}
\affiliation{Research Center for Functional Materials, National Institute for Materials Science (NIMS), Tsukuba, Ibaraki 305-0044, Japan}

\author{Tie-Fu Li}
\thanks{litf@tsinghua.edu.cn}
\affiliation{Institute of Microelectronics and Frontier Science Center for Quantum Information, Tsinghua University, Beijing 100084, China}
\affiliation{Beijing Academy of Quantum Information Sciences, Beijing 100193, China}

\author{Limin Tong}
\affiliation{State Key Laboratory of Modern Optical Instrumentation, College of Optical Science and Engineering, Zhejiang University, Hangzhou 310027, China}

\author{Shi-Yao Zhu}
\affiliation{Interdisciplinary Center of Quantum Information, State Key Laboratory of Modern Optical Instrumentation, and Zhejiang Province Key Laboratory of Quantum Technology and Device, Department of Physics, Zhejiang University, Hangzhou 310027, China}

\author{Franco Nori}
\affiliation{Theoretical Quantum Physics Laboratory, RIKEN Cluster for Pioneering Research, Wako-shi, Saitama 351-0198, Japan}
\affiliation{Physics Department, University of Michigan, Ann Arbor, Michigan 48109-1040, USA}

\author{J. Q. You}
\thanks{jqyou@zju.edu.cn}%
\affiliation{Interdisciplinary Center of Quantum Information, State Key Laboratory of Modern Optical Instrumentation, and Zhejiang Province Key Laboratory of Quantum Technology and Device, Department of Physics, Zhejiang University, Hangzhou 310027, China}

\date{\today}

\begin{abstract}
Exceptional points (EPs) are exotic degeneracies of non-Hermitian systems, where the eigenvalues and the corresponding eigenvectors simultaneously coalesce in parameter space, and these degeneracies are sensitive to tiny perturbations on the system. Here we report an experimental observation of the EP in a hybrid quantum system consisting of dense nitrogen (P1) centers in diamond coupled to a coplanar-waveguide resonator. These P1 centers can be divided into three subensembles of spins, and cross relaxation occurs among them. As a new method to demonstrate this EP, we pump a given spin subensemble with a drive field to tune the magnon-photon coupling in a wide range. We observe the EP in the middle spin subensemble coupled to the resonator mode, irrespective of which spin subensemble is actually driven. This robustness of the EP against pumping reveals the key role of the cross relaxation in P1 centers. It offers a novel way to convincingly prove the existence of the cross-relaxation effect via the EP.
\end{abstract}

\maketitle


\section{Introduction}
An exceptional point (EP) corresponds to the singularity of a non-Hermitian Hamiltonian, where the eigenvalues coalesce together (see, e.g., Refs.~\cite{EP1,El-Ganainy-2018,Ozdemir-2019}). Meanwhile, the corresponding eigenvectors also coincide at this point, as guaranteed by the non-Hermiticity of the Hamiltonian~\cite{EP1}. The presence of the EP can give rise to many exotic physical phenomena, such as the revival of lasing~\cite{Brand,pengb}, nonreciprocal energy transfer~\cite{hu}, directional lasing~\cite{pengb2}, asymmetric mode switching~\cite{JDO}, controllable coherence in lasing systems~\cite{PhysRevLett.108.173901}, and sensitivity enhancement of detection~\cite{PRL203901,PRL110802,nature548187,nature548192,Cao19,Wu20,Khurgin20}. Experimentally, EPs have been found in various physical systems, including coupled optical cavities~\cite{naturephys}, coupled microwave resonators~\cite{PhysRevLett.86.787}, atom-cavity composites~\cite{PhysRevLett.104.153601}, exciton-polariton systems~\cite{nature526}, magnonic systems~\cite{zhangdengke,xufeng,Du-PRAppl}, coupled acoustic cavities~\cite{Chan16}, and coupled ferromagnetic waveguides~\cite{Chan18}.
Owing to the good coherence of a spin ensemble and its strong coupling to a cavity, hybrid spin ensemble-cavity systems have promising applications in quantum information processing~\cite{Xiang13,Kurizki15,Nakamura19} and offer a platform to demonstrate intriguing phenomena, e.g., the bistability of magnon polaritons~\cite{Wang18}, gradient memory~\cite{Zhang15-1}, nonreciprocity~\cite{Nakamura-PRL16,Wang19}, level attraction~\cite{Harder18,Grigoryan18}, and magnon-related quantum entanglement~\cite{Agarwal18,Agarwal19,Nakamura-Sci20}.

For a small yttrium-iron-garnet (YIG) sphere in a three-dimensional cavity~\cite{Nakamura19}, one can vary the magnon-photon coupling by moving the sphere in the cavity. However, this mechanical tunability is not precise, which limits its applications in quantum technologies. Also, Ref.~\cite{Potts20} has proposed to control the interaction between magnons and photons via a tunable microwave cavity. Actually, there are various hybrid systems with a spin ensemble fixed on a coplanar-waveguide resonator~\cite{PhysRevLett.107.220501,PhysRevLett.105.140501,Angerere1701626,
PhysRevLett.105.140502, PhysRevLett.110.067004,PhysRevLett.123.107701,PhysRevLett.123.107702}. For these on-chip systems, one also needs to explore a method to precisely tune the magnon-photon coupling.
In this work, we experimentally investigate a hybrid quantum system consisting of dense nitrogen (P1) centers in diamond coupled to a coplanar-waveguide resonator.
Different from the ferromagnetic spin ensemble in a YIG sphere~\cite{Huebl13,Tabuchi14,Zhang14,Bai15,Zhang15},
the P1 centers constitute a {\it paramagnetic} spin ensemble. We find that we can excite a large number of magnons with a drive field to precisely tune the magnon-photon coupling in a wide range and make it feasible to demonstrate the EP in this on-chip hybrid system. The term \emph{magnon} is often used to describe collective excitations in a magnetically ordered system (e.g., a ferromagnetic system) with strong spin-exchange interactions. In the long-wavelength limit, the effect of the exchange interactions can be ignored and the dipolar interactions become dominant~\cite{White07}. This limiting case is analogous to paramagnetic systems where the dipolar interactions play the main role. In those paramagnetic systems, magnons can also be used to characterize the collective spin excitations (see, e.g., Refs.~\cite{Li18,Wesenberg09,Wu10}). In the same manner, here we characterize the collective excitations of P1 centers by harnessing magnons.

The P1 centers in diamond can be divided into three spin subensembles and cross relaxation
can occur among them~\cite{PhysRevLett.105.140501, PhysRevLett.110.067004,PhysRevB.91.140408,PhysRev.114.445,Sorokin60,Breeze18}.
With the magnon frequency of the middle subensemble tuned in resonance with the resonator mode, we observe an EP by driving {\it any} of the three spin subensembles with a microwave field. Related to the middle subensemble, the other two subensembles are in the dispersive regime and each produces a frequency shift to the resonator mode. These two frequency shifts depend on the magnon occupations and can cancel each other when the magnon occupations in the other two subensembles become equal. Here the EP is observed by pumping any of the three spin subensembles, indicating that the same magnon occupation is achieved for each subensemble, due to the cross-relaxation effect.

Also, it is known that an EP can enhance the sensitivity to a perturbation~\cite{PRL203901,PRL110802,nature548187,nature548192,Cao19,Wu20,Khurgin20}, while our observed transmission spectrum does not show any appreciable differences at the EP when pumping any of the three spin subensembles. This robustness of the EP against driving reveals the key role of the cross relaxation, which induces the same magnon occupation in each spin subensemble. To the best of our knowledge, this is the first experimental observation of the EP in this hybrid quantum system and it could potentially push the field of these hybrid quantum systems in a new direction. Moreover, it also offers a novel way of convincingly proving the existence of the cross-relaxation effect via the EP. In contrast to the cavity magnonics system consisting of a ferromagnetic spin ensemble coupled to a three-dimensional microwave cavity~\cite{zhangdengke,xufeng,Du-PRAppl}, this on-chip hybrid system has a more compact configuration and significantly reduces the system size.

\section{Exceptional point of the hybrid quantum system}

\begin{figure}[!hbt]
\centering
\includegraphics[scale=0.5]{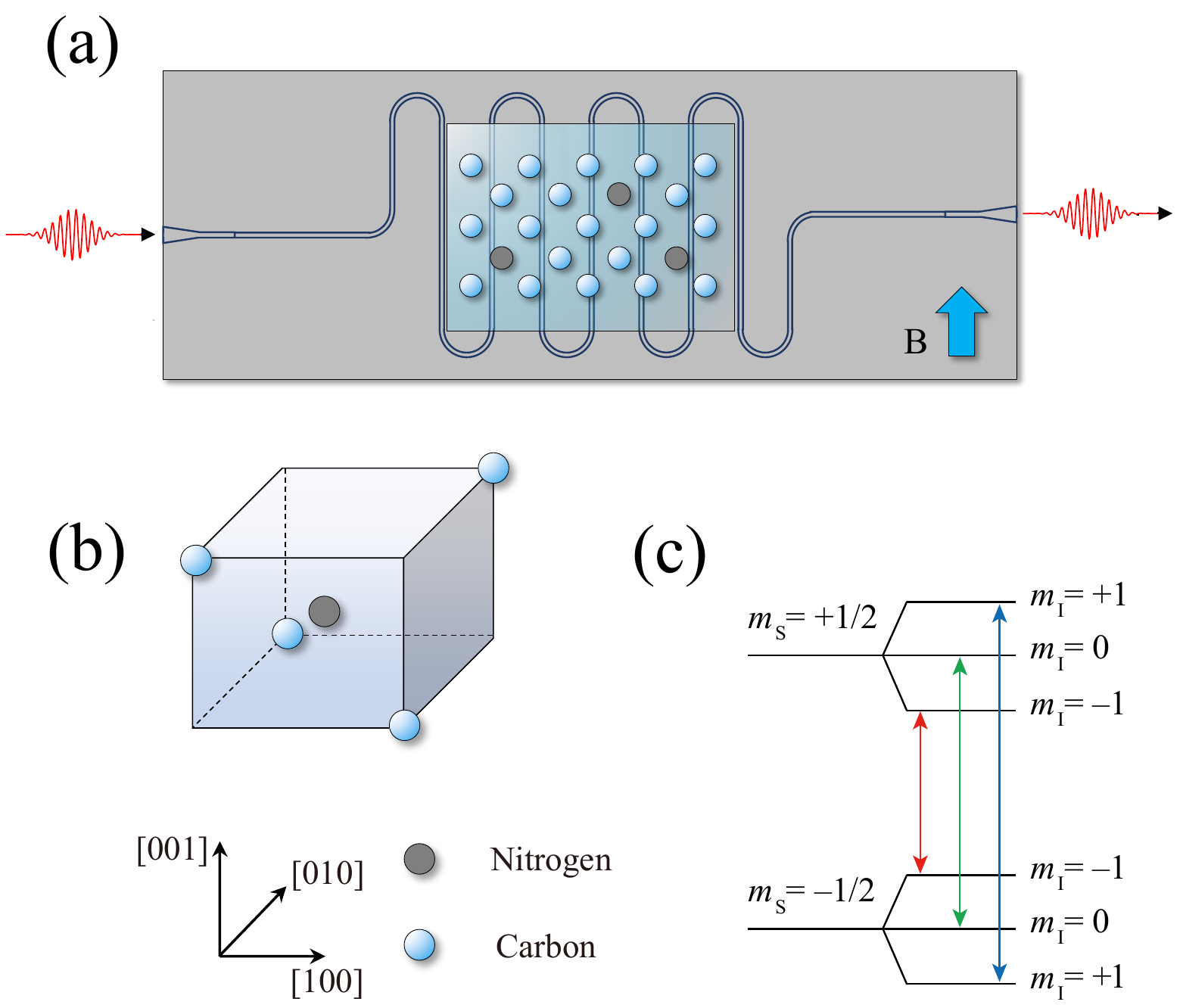}
\caption{(a) A schematic of the hybrid system. The diamond sample with P1 centers is glued on a coplanar-waveguide resonator. (b) A P1 center that involves the substitution of a carbon atom by a nitrogen atom.
(c) The energy levels and allowed state transitions in the P1 center.}
\label{fig-sample}
\end{figure}

The hybrid system that we study is composed of P1 centers in diamond glued on a coplanar-waveguide resonator (see Fig.~\ref{fig-sample} and Appendix \ref{appendix-A}). The [001] crystal axis of the diamond is perpendicular to the surface of the resonator and the static magnetic field $B$ is applied along the [100] crystal axis. The P1 centers, which are the main defects in type-1b diamond synthesized under both high pressure and high temperature, are formed by substituting some carbon atoms with nitrogen atoms. These P1 centers can be divided into three subensembles of spins ($s=0,\pm$) with transition frequencies $\omega_0=\gamma_eB$ and $\omega_{\pm}=\gamma_eB\pm A_{\parallel}$, where  $\gamma_e/2\pi = 28$~GHz/T is the gyromagnetic ratio and $A_{\parallel}/2\pi = 94$~MHz is due to the hyperfine interaction in P1 centers.

\begin{figure*}[!hbt]
\flushleft
\includegraphics[scale=0.35]{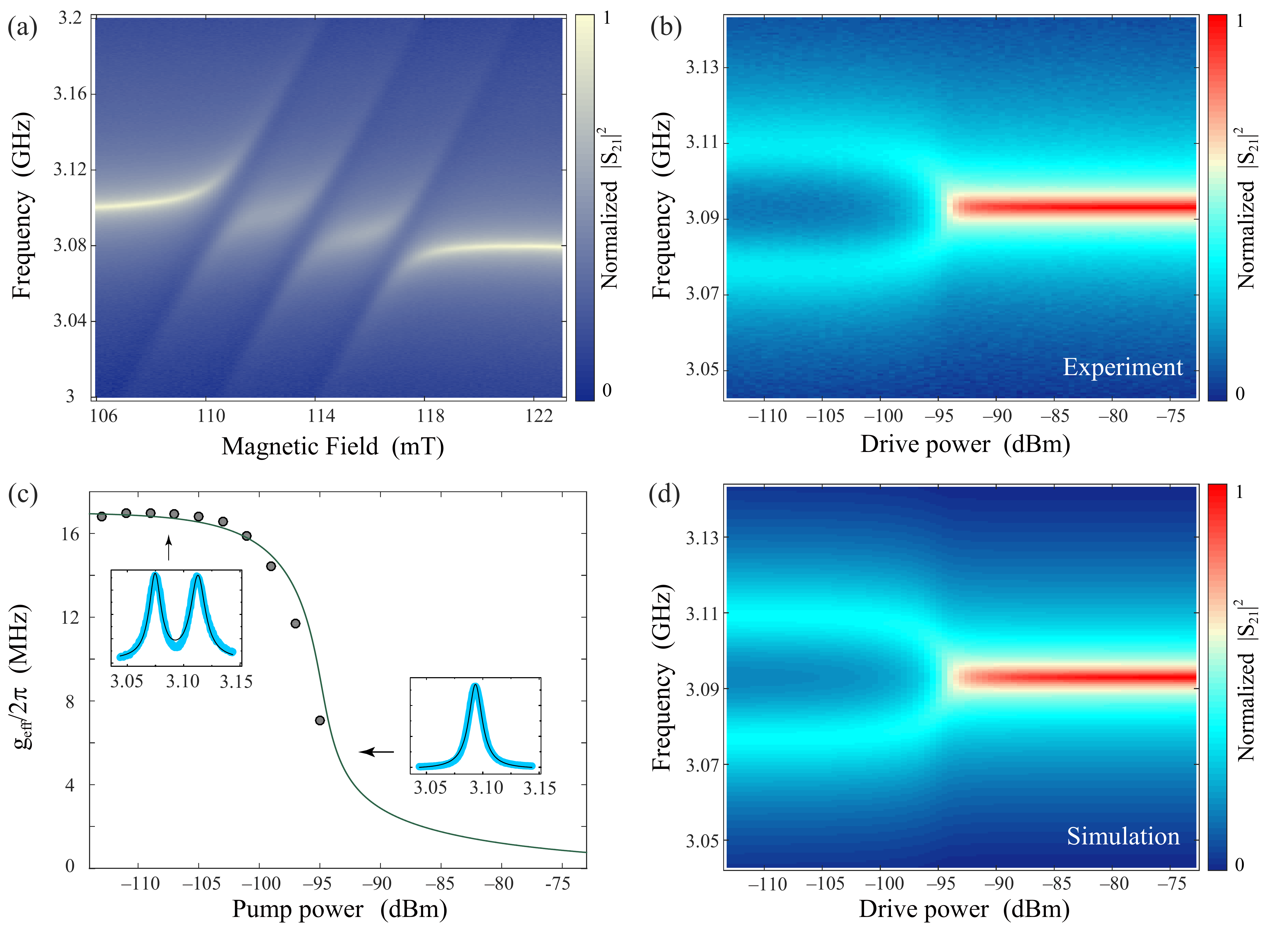}
\caption{The experimental realization of the exceptional point. (a) The transmission spectrum measured with a probe tone of power $-120$~dBm, corresponding to an average photon number $\overline{n} = 404 $. (b) The Rabi splitting due to the coupling between the magnons of the $s = 0$ subensemble and the resonator photons versus the power of the drive tone. Here, the static magnetic field is tuned to have the magnons of the $s = 0$ subensemble in resonance with the resonator mode: $\omega_{0}/2\pi=\omega_{c}/2\pi=$3.093~GHz. Note that the decay rate of the resonator is nearly independent of the power of the drive tone [cf. Fig.~\ref{fig-frequency}(b) in Appendix \ref{appendix-C}]. (c)~The effective magnon-photon coupling (gray circles) extracted from the Rabi splittings in (b). The fitting curve is obtained using Eq.~(\ref{driving}), with parameters $g/2\pi=$ 17.2~MHz, $\kappa/2\pi=0.6$~MHz, and $\gamma/2\pi=11.9$~MHz. (d) The transmission spectrum simulated with Eq.~(\ref{s21}), where the fitting curve in (c) is used for the effective magnon-photon coupling.}
\label{fig-EP}
\end{figure*}

The collective spin excitations in each subensemble are magnons. As shown in the measured transmission spectrum [Fig.~\ref{fig-EP}(a)], there are three appreciable anticrossings, which reveal the formation of magnon polaritons by the strong coupling between the magnons of $s=0,\pm$ subensembles and the photons in the superconducting coplanar-waveguide resonator.

For clarity, we first focus on the interaction between the magnons in the $s=0$ spin subensemble and the photons in the resonator. The model Hamiltonian can be written as (see Appendix \ref{appendix-B})
$H=\omega_{c}a^{\dag}a+\omega_{0}b^{\dagger}b
      +g_{\rm eff}(a^{\dagger}b+ab^{\dagger})$,
where $a$ and $b$ are annihilation operators of the photons and magnons, with frequencies $\omega_c$ and $\omega_0$, respectively. In ferromagnetic materials, spins are very polarized and less magnons are excited, so the effective magnon-photon coupling is often approximated as a constant~\cite{Tabuchi14,Zhang14,Zhang15,Bai15} $g_{\rm eff}\approx g$. Here, the P1 centers are paramagnetic and considerable numbers of magnons can be easily excited. Therefore, we need to consider the dependence of the magnon-photon coupling on the occupation number of magnons, $g_{\rm eff}=g\sqrt{1-\langle b^{\dag}b\rangle/(N/2)}$, where $N$ is the number of P1 centers in the sample.

Using the input-output theory~\cite{Walls94}, we can derive the transmission amplitude of the hybrid system,
\begin{equation}
\label{s21}
S_{21} =   \frac{2\sqrt{\kappa_{i}\kappa_{o}}}{\kappa+i(\omega_{c}-\omega)+\frac{g_{\rm eff}^{2}}{\gamma+i(\omega_{0}-\omega)}},
\end{equation}
where $\gamma$ is the damping rate of the magnon mode, $\kappa_{i(o)}$ is the decay rate of the resonator mode due to the input (output) port, and $\kappa=\kappa_i+\kappa_o+\kappa_{\rm int}$, with $\kappa_{\rm int}$ being the intrinsic decay rate of the resonator. The damping of the magnon mode may result from the dipolar interactions of the P1 center with neighboring $^{13}$C nuclei and other P1 centers~\cite{Wyk97} as well as the inhomogeneities of the external static magnetic field and the magnetic field of the resonator mode. For our coplanar-waveguide resonator, $\kappa_{i}\approx\kappa_{o}\gg\kappa_{\rm int}$. Away from the left or right anticrossing point, the polariton mode approaches the resonator mode and its line width gives $\kappa/2\pi=0.6\pm 0.05$~MHz. When measuring the transmission spectrum in Fig.~\ref{fig-EP}(a), we do not apply a drive tone to the system. Thus, $\langle b^{\dag}b\rangle\approx 0$ at cryogenic temperatures. We obtain $g_{\rm eff}/2\pi\approx g/2\pi=17.2\pm 0.5$~MHz from the Rabi splitting at the middle anticrossing point. Finally, we use Eq.~(1) to fit the measured transmission spectrum around the middle anticrossing point, which gives $\gamma/2\pi=11.9\pm 0.3$~MHz.

When including both the decay rate of the resonator mode and the damping rate of the magnon mode~\cite{Nori19,Nori20}, the Hamiltonian of the hybrid system can be effectively written as a non-Hermitian Hamiltonian:
\begin{equation}\label{matrix}
H_{\rm eff}=(\omega_{c}-i\kappa)a^{\dag}a+(\omega_{0}-i\gamma)b^{\dagger}b+g_{\rm eff}(a^{\dagger}b+ab^{\dagger}).
\end{equation}
In matrix form, we can write the effective Hamiltonian as
\begin{equation}
H_{\rm eff}=\begin{pmatrix}
  \omega_{c}-i\kappa & g_{\rm eff}  \\
  g_{\rm eff} & \omega_{0}-i\gamma  \\
 \end{pmatrix},
\end{equation}
which has two complex eigenvalues,
\begin{equation}
\label{eigenvalues}
\omega_{1,2}=\frac{1}{2}\left[\omega_{c}+\omega_{0}-i(\kappa+\gamma)\pm\sqrt{4g_{\rm eff}^2-(\gamma-\kappa+i\omega_{0}-i\omega_{c})^2}\right].
\end{equation}
When the magnon is tuned in resonance with the resonator mode ($\omega_{c}=\omega_0$), the two eigenvalues of the magnon polaritons are reduced to
\begin{equation}
\label{EQ3}
\omega_{1,2}=\omega_{0}-\frac{1}{2}i(\kappa+\gamma)\pm\frac{1}{2}\sqrt{4g_{\rm eff}^2-(\gamma-\kappa)^2}.
\end{equation}
For $2g_{\rm eff}>(\gamma-\kappa)$, Eq.~(\ref{EQ3}) gives two separate polariton modes at the anticrossing point, each with line width $\frac{1}{2}(\kappa+\gamma)$. However, these two polariton modes coalesce to one at
$2g_{\rm eff}=(\gamma-\kappa)$, which is the EP of the hybrid system. When $2g_{\rm eff}<(\gamma-\kappa)$,
$\omega_{1,2}=\omega_{0}-\frac{1}{2}i(\kappa+\gamma)\pm \frac{1}{2} i\sqrt{(\gamma-\kappa)^2-4g_{\rm eff}^2}$,
indicating that the two polariton modes have different line widths.
Corresponding to the two eigenvalues in Eq.~(\ref{EQ3}), the two eigenvectors also coalesce at the EP and each behaves differently on the two sides of the EP (cf.~Appendix \ref{appendix-B}).

To demonstrate the EP, we tune $g_{\rm eff}$ via $\langle b^{\dag}b\rangle$ by applying a drive tone of frequency $\omega_{d}/2\pi= 3.093$~GHz to the system, which is in resonance with the magnons of the $s=0$ subensemble. To guarantee that the system is in its stationary state for each measurement, this driving tone lasts for 3000~s before the next measurement is performed (cf.~Appendix \ref{appendix-C}). Figure~\ref{fig-EP}(b) shows the measured transmission spectrum of the system versus the power of the drive tone, where the static magnetic field is fixed at the middle anticrossing point in Fig.~\ref{fig-EP}(a). This demonstrates the driving-power dependence of the Rabi splitting for the two polariton branches. It is clear that the EP occurs at $P_d\approx -93.7$~dBm. In the region of $P_d< -93.7$~dBm, the two polariton branches are separated but when increasing $P_d$, the separation (i.e., the Rabi splitting $2g_{\rm eff}$) decreases and, at the EP, the two peaks coalesce to one. We extract these data from the Rabi splitting and show the behavior of $g_{\rm eff}$ versus the drive power $P_d$ in Fig.~\ref{fig-EP}(c) (cf. the gray circles). There is a relation between the reduced magnon occupation $\chi\equiv\langle b^{\dag}b\rangle/N$ and the Rabi frequency $\Omega_d$ ($\propto \sqrt{P_d}$) of the drive tone (see Appendix \ref{appendix-D}),
\begin{equation}
\label{driving}
(\gamma+\eta\kappa)^{2}\chi-\xi\eta(\Omega_{d}/\sqrt{N})^2=0,
\end{equation}
where $\xi=(1-2\chi)/(1-\chi)$ and $\eta=g^2_{\rm eff}/\kappa^2$. We use this relation to fit the data of $g_{\rm eff}$ [see the curve in Fig.~\ref{fig-EP}(c)] and obtain $\Omega_d=k\sqrt{P_d}$, with the fitting parameter $k=1.54\times 10^{14}$. The theoretical value of $k$ is $k= \sqrt{\kappa/2\hbar \omega_{c}}= 9.59 \times 10^{14}$, which is larger than the above fitting value. This is due to the leakage of the drive power in the setup. Next, we use the fitted results [i.e., the curve in Fig.~\ref{fig-EP}(c)] and Eq.~(\ref{s21}) to simulate the transmission spectrum [Fig.~\ref{fig-EP}(d)]. Clearly, this simulated transmission spectrum agrees well with the experimental results in Fig.~\ref{fig-EP}(b).

\begin{figure}[!hbt]
\flushleft
\includegraphics[scale=0.37]{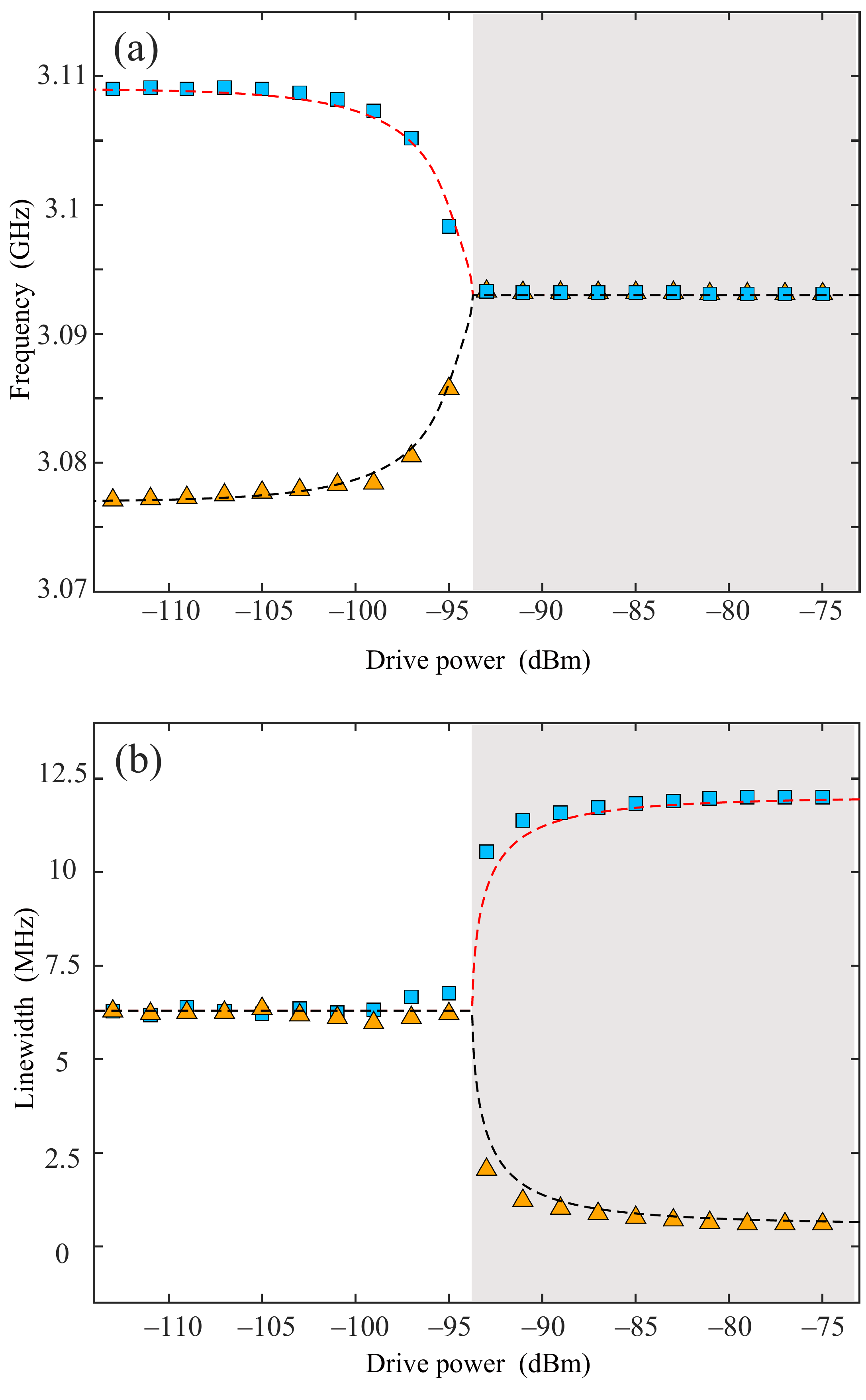}
\caption{The peak positions and line widths of the two polariton modes. (a)~The data shown by squares and triangles are the peak positions of the two polariton modes extracted from Fig.~\ref{fig-EP}(b) and the solid curves are the real parts of the two eigenvalues calculated using Eq.~(\ref{EQ3}). (b)~The data denoted by squares and triangles are the line widths of the two polariton modes extracted from Fig.~\ref{fig-EP}(b) and the solid curves are the imaginary parts of the two eigenvalues calculated using Eq.~(\ref{EQ3}). When $P_d >93.7$~dBm, only one peak is measured. We use both Eq.~(\ref{s21}) and Eq.~(\ref{EQ3}) to deduce the line widths of the two modes in this region.}
\label{fig-positions}
\end{figure}

To make the observation of the EP more convincing, we compare the experimental data with the real and imaginary parts of $\omega_{1,2}$ in Eq.~(\ref{EQ3}) by varying the drive power $P_d$. The two curves in Figs.~\ref{fig-positions}(a) and \ref{fig-positions}(b) correspond to the real and imaginary parts of $\omega_{1,2}$ in Eq.~(\ref{EQ3}), where the dependence of the effective magnon-photon coupling $g_{\rm eff}$ on the drive power is given by the fitting curve in Fig.~\ref{fig-EP}(c). These two curves show the characteristics of the EP. In Fig.~\ref{fig-positions}(a), the experimental data are extracted from the peak positions of the two polariton branches in Fig.~\ref{fig-EP}(b); and the experimental data in Fig.~\ref{fig-positions}(b) correspond to the line widths of the two magnon polaritons. At $P_d\approx 93.7$~dBm, the two polariton modes coalesce and only one peak is visible in the region of $P_d >93.7$~dBm. To obtain the line widths of the two modes in this region, we first fit the transmission spectrum and acquire the effective coupling strength $g_{\rm{eff}}$.  Then, we use Eq.~(\ref{EQ3}) and the fitted $g_{\rm{eff}}$ to infer the line widths of the two modes, as in Ref.~\cite{naturephys}. Indeed, the experimental data show a good agreement with the theoretical results, further confirming the observation of the EP in our hybrid system.

\section{Cross-relaxation effect in P1 centers}\label{cross-relaxation-effect}

Above, we only consider the coupling between the $s=0$ spin subensemble and the resonator mode. In fact, as shown in Fig.~\ref{fig-EP}(a), magnons in other two spin subensembles ($s=\pm$) are also strongly coupled to the resonator mode. The results in Fig.~\ref{fig-EP}(a) show that the magnon-photon coupling strengths are nearly equal for these three subensembles. Here, the static magnetic field is tuned to have the magnons of the $s=0$ subensemble in resonance with the resonator mode, so the magnons in other two spin subensembles are very off resonant with the resonator mode. In such a {\it dispersive} regime, each of the $s=\pm$ subensembles yields a frequency shift to the resonator mode (Appendix \ref{appendix-E}) and the frequency of the resonator mode is shifted from $\omega_c$ to
\begin{equation}
\tilde{\omega}_{c}=\omega_{c}+\frac{g^2_{\rm eff,+}}{\delta_{+}}+\frac{g^2_{\rm eff,-}}{\delta_{-}},
\end{equation}
where $g_{\rm eff,\pm}$ and $\delta_{\pm}=\omega_c-\omega_{\pm}$ are the effective magnon-photon coupling strengths and frequency detunings between the resonator mode and the $s=\pm$ subensembles, respectively. In Fig.~\ref{fig-EP}(b), the frequency of the drive tone is $\omega_{d}/2\pi= 3.093$~GHz, which is in resonance with the magnons of the $s=0$ subensemble. Owing to the cross relaxation in P1 centers~\cite{PhysRev.114.445,Sorokin60}, identical occupations of magnons can be induced in other two spin ensembles when the drive tone is applied for a long time~\cite{PhysRevLett.110.067004}. Thus, $g_{\rm eff,+}=g_{\rm eff,-}$. If this is obeyed when varying the drive power, due to $\delta_{+}=-\delta_{-}$, the frequency shifts of the resonator mode induced by the $s=\pm$ subensembles then cancel each other and $\tilde{\omega}_{c}$ is reduced to $\omega_c$. Therefore, the frequency of the resonator mode is not modified by the presence of the $s=\pm$ spin subensembles, owing to the cross-relaxation effect that induces $g_{\rm eff,+}=g_{\rm eff,-}$. Indeed, the observation of the EP in Fig.~\ref{fig-EP}(b) reveals that the frequency of the resonator mode does not vary when tuning the drive power. Otherwise, the resonant condition $\tilde{\omega}_{c}=\omega_0$ for observing the EP cannot be satisfied.

\begin{figure*}[!hbt]
\flushleft
\includegraphics[scale=0.36]{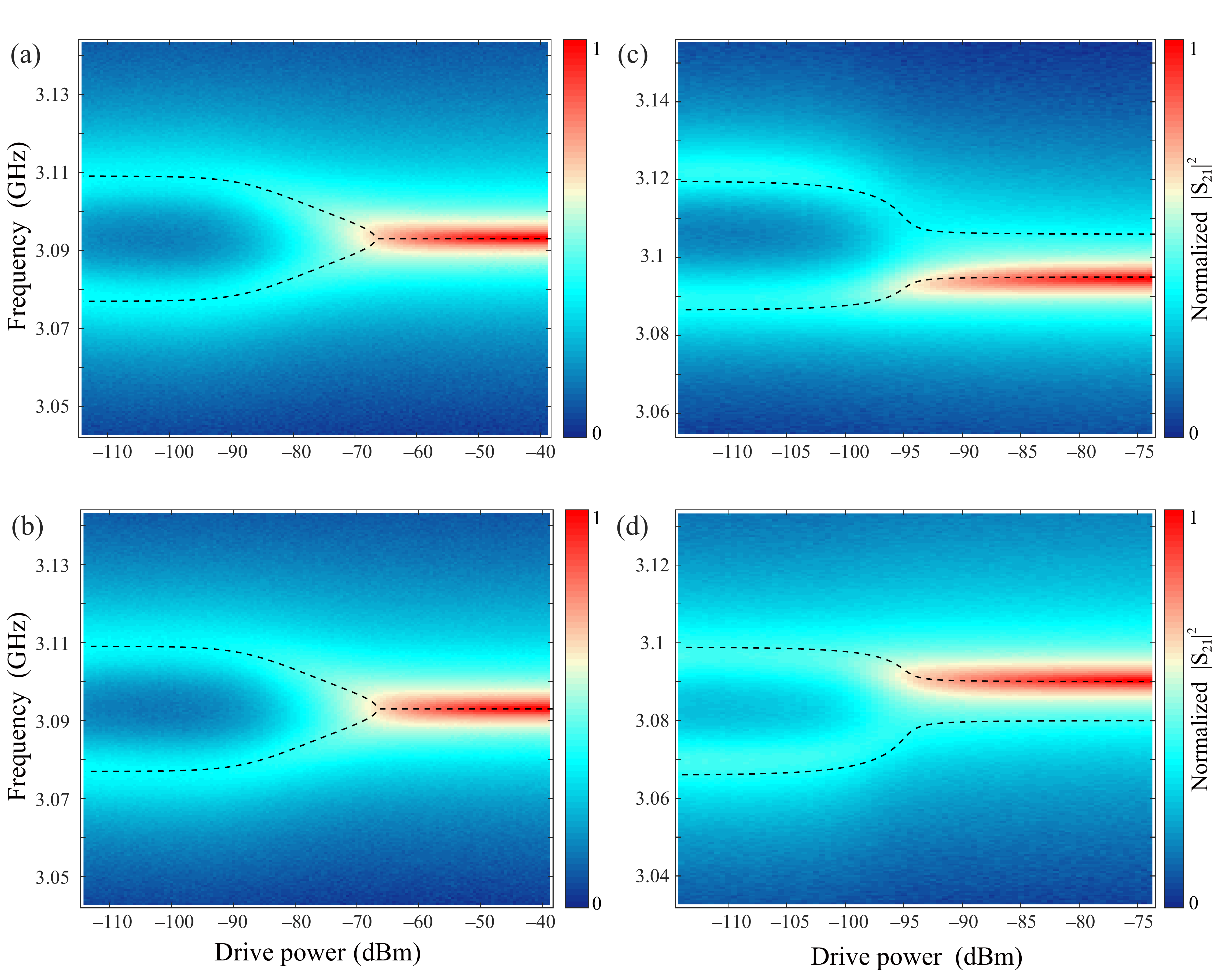}
\caption{The effect of the cross relaxation. (a),(b) The Rabi splittings related to the $s=0$ subensemble, where the drive tone is applied in resonance with the magnons of the $s=+$ and $-$ subensembles, respectively. The dashed curves are simulated as the real part of Eq.~(\ref{EQ3}). (c),(d) The Rabi splittings related to the $s=+$ and $-$ subensembles, where the drive tone is in resonance with the magnons of the $s=+$ and $-$ subensembles, respectively. The dashed curves are simulated as the real part of Eq.~(\ref{eigenvalues}), but with $\omega_c$, $\omega_0$ and $g_{\rm{eff}}$ replaced by $\tilde{\omega}_{c}$, $\omega_\pm$ and $g_{\rm{eff}, \pm}$ respectively. In (c), the magnons of the $s=+$ subensemble are tuned to have frequency $\omega_+/2\pi=$  3.106 GHz via the static magnetic field. The magnons of the $s=0$ and $-$ subensembles have frequencies $\omega_0/2\pi=$ 3.012 GHz and $\omega_-/2\pi=$ 2.918 GHz. The frequency of the resonator mode is found to be $\omega_c/2\pi=$ 3.095 GHz, which is slightly blue shifted from $\omega_c/2\pi=$ 3.093 GHz in Fig.~\ref{fig-EP} due to the static magnetic field (see Appendix \ref{appendix-C}). In (d), the magnons of the $s=-$ subensemble are tuned to have frequency $\omega_-/2\pi=$ 3.080 GHz via the static magnetic field. The magnons of the $s=0$ and $+$ subensembles have frequencies $\omega_0/2\pi=$ 3.175 GHz and $\omega_+/2\pi=$ 3.269 GHz. The frequency of the resonator mode is found to be $\omega_c/2\pi=$ 3.090 GHz, which is slightly red shifted from $\omega_c/2\pi=$ 3.093 GHz in Fig.~\ref{fig-EP} due to the static magnetic field. The other parameters used are the same as in Fig.~\ref{fig-EP}.}
\label{fig-cross}
\end{figure*}

While the static magnetic field is tuned to have the magnons of the $s=0$ subensemble in resonance with the resonator mode, we also manage to measure the transmission spectrum of the hybrid system by resonantly pumping the magnons in the $s=+$ and $s=-$ subensembles, respectively [Figs.~\ref{fig-cross}(a) and \ref{fig-cross}(b)], instead of the magnons in the $s=0$ subensemble. The results reveal that the transmission spectrum is also symmetric about the resonator frequency $\omega_c$ and the EP occurs as well, as in Fig.~\ref{fig-EP}(b).
This means that $g_{\rm eff,+}=g_{\rm eff,-}$, irrespective of which spin subensemble is pumped by the drive tone. This further proves the equal occupations of magnons in the $s=\pm$ subensembles and demonstrates the robustness of the EP against driving in this hybrid system. In Figs.~\ref{fig-cross}(a) and \ref{fig-cross}(b), the simulated peak positions of the two polariton modes are shown, which also match the experimental results. Compared to the position of the EP in Fig.~\ref{fig-EP}(b), the EP in Figs.~\ref{fig-cross}(a) and \ref{fig-cross}(b) shifts to a higher drive power. This is because the drive tone now has a frequency detuning from the resonator mode and the relation between the magnon occupation and the drive power becomes different from Eq.~(\ref{driving}) (cf. Appendix \ref{appendix-D}).

\section{Discussion and Conclusions}

We show how the EP can be removed by the varying frequency shift of the resonator mode when changing the drive power.
The static magnetic field is tuned to have the magnons of the $s=+$ spin subensemble nearly resonant with the resonator mode.  Then, we resonantly pump the magnons in this subensemble with a drive tone. Figure~\ref{fig-cross}(c) shows the measured transmission spectrum of the hybrid system versus the drive power. As in Fig.~\ref{fig-EP}(b), this applied drive tone lasts for 3000~s before the next measurement is performed. The measured transmission spectrum becomes asymmetric and no EP is observed, in sharp contrast to Fig.~\ref{fig-EP}(b). In the present case, the magnons of the $s=0,-$ subensembles are in the dispersive regime related to the resonator mode, yielding that the frequency of the resonator mode shifts from $\omega_c$ to $\tilde{\omega}_{c}=\omega_{c}+g^2_{\rm eff}/\delta_{0}+g^2_{\rm eff,-}/\delta_{-}$, where the frequency detunings are $\delta_{0}=\omega_c-\omega_{0}$ and $\delta_{-}=\omega_c-\omega_{-}$. The frequencies $\omega_{1,2}$ of the two polariton modes are also given by Eq.~(\ref{eigenvalues}) but with $\omega_c$, $\omega_0$ and $g_{\rm eff}$ replaced by $\tilde{\omega}_{c}$, $\omega_{+}$ and $g_{{\rm eff},+}$, respectively. Now, the terms $(g^2_{\rm eff}/\delta_{0}+g^2_{\rm eff,-}/\delta_{-})$ in $\tilde{\omega}_{c}$ vary with the drive power, so the resonant condition $\tilde{\omega}_{c}=\omega_{+}$ cannot always be obeyed. Therefore, we cannot reduce Eq.~(\ref{eigenvalues}) to the simple form of Eq.~(\ref{EQ3}) to exhibit the EP.

In Fig.~\ref{fig-cross}(d), we also measure the transmission spectrum of the hybrid system versus the drive power but we tune the static magnetic field to have the magnons of the $s=-$ subensemble nearly resonant with the resonator mode and then resonantly pump the magnons in this subensemble with a drive tone. As in Fig.~\ref{fig-cross}(c), the transmission spectrum is asymmetric and no EP is observed. In this case, the magnons of the $s=0,+$ subensembles are in the dispersive regime related to the resonator mode and the frequency of the resonator mode is shifted from $\omega_c$ to $\tilde{\omega}_{c}=\omega_{c}+g^2_{\rm eff}/\delta_{0}+g^2_{\rm eff,+}/\delta_{+}$, where $\delta_{0}=\omega_c-\omega_{0}$ and $\delta_{+}=\omega_c-\omega_{+}$. Also, this modified frequency of the resonator mode varies with the drive power, so that the resonant condition $\tilde{\omega}_{c}=\omega_{-}$ cannot always be satisfied. The frequencies $\omega_{1,2}$ of the two polariton branches are given by Eq.~(\ref{eigenvalues}) as well but with $\omega_c$, $\omega_0$, and $g_{\rm eff}$ replaced by $\tilde{\omega}_{c}$, $\omega_{-}$, and $g_{{\rm eff},-}$, respectively. Compared to Fig.~\ref{fig-EP}(b), the EP is removed in Figs.~\ref{fig-cross}(c) and \ref{fig-cross}(d), due to the varying frequency shift of the resonator mode with the drive power.

Some experimental results on the drive-power dependence of the effective coupling have been presented in Refs.~\cite{PhysRevLett.110.067004,Weichselbaumer19} but no EP has been observed there. In Ref.~\cite{PhysRevLett.110.067004},
the Rabi splitting versus the drive power has been shown only for the $s=+$ subensemble. As discussed above, in our study the EP does not occur in this case [cf. Fig.~\ref{fig-cross}(c)], because the resonant condition $\tilde{\omega}_{c}=\omega_{+}$ cannot always be obeyed for the $s=+$ subensemble. Also, as analyzed in Sec.~\ref{cross-relaxation-effect}, only for the $s=0$ subensemble can the EP occur in the hybrid quantum system considered. To confirm the observation of the EP, it is essential to both show the experimental results in Fig.~\ref{fig-EP}(b) for the $s=0$ subensemble and demonstrate the coalescence of the peak positions and line widths of the two polariton modes at the EP (i.e., Fig.~\ref{fig-positions}). No such results have been shown in Refs.~\cite{PhysRevLett.110.067004,Weichselbaumer19}.

The results in Refs.~\cite{PhysRevLett.110.067004,Weichselbaumer19} have been explained by a depolarization model involving the mixed states of the system. In our theory, the system is described by a non-Hermitian Hamiltonian $H_{\rm eff}$ in Eq.~(\ref{matrix}). Then, the density operator $\rho(t)$ of the system is governed by the following Liouvillian equation~\cite{Nori19}:~$\partial \rho(t)/\partial t=-i[H_{\rm eff}\rho(t)-\rho(t)H_{\rm eff}^{\dag}]$. Thus, $\rho(t)$ can be explicitly expressed as $\rho(t)=\exp(-iH_{\rm eff}t)\rho(0)\exp(iH_{\rm eff}^{\dag}t)$. For pure states of the system, $\rho^2(t)=\rho(t)$, while $\rho^2(t)\neq \rho(t)$ in our case (see Appendix~\ref{appendix-B}). This reveals that the system is in the {\it mixed} state, even if the non-Hermitian Hamiltonian of the system has the simple form in Eq.~(\ref{matrix}). Therefore, the conclusion regarding the mixed states of the system is consistent with the depolarization model in Refs.~\cite{PhysRevLett.110.067004,Weichselbaumer19}. Our model is based on the concepts and methods of non-Hermitian physics.
Compared with the depolarization model, our model has the following three distinct merits. First, using the Holstein-Primakoff transformation and the mean-field approximation, we show that the effective coupling $g_{\rm eff} = g\sqrt{1-\langle b^{\dag}b\rangle/(N/2)}$ in our model is related to the magnon occupation $\langle b^{\dag}b\rangle$, where $\langle b^{\dag}b\rangle$ can be obtained using the quantum Langevin equation. By defining the spin-polarization factor in Refs.~\cite{PhysRevLett.110.067004,Weichselbaumer19} as $P_{\rm eff}=1-\langle b^{\dag}b\rangle/(N/2)$, the effective coupling strength in our model can be converted to $g_{\rm eff} = g\sqrt{P_{\rm eff}}$ in the depolarization model. Thus, our model provides a microscopic interpretation of the spin-polarization factor. Second, in our model we derive the effective non-Hermitian Hamiltonian of the hybrid system and convincingly explain the experimental results related to the EP by analyzing the real and imaginary parts of the eigenvalues of the effective non-Hermitian Hamiltonian, which is beyond the depolarization model. In contrast, the depolarization model in Refs.~\cite{PhysRevLett.110.067004,Weichselbaumer19} only gives the relation between the effective coupling and the drive power but does not provide the effective non-Hermitian Hamiltonian and its eigenvalues (i.e., it cannot be used to describe the EP). Third, while our model uses the framework of the effective non-Hermitian Hamiltonian, which is different from the depolarization model, it may also be harnessed to study other phenomena of non-Hermitian physics~\cite{Ashida20}. Thus, our study sheds new light on power-induced depolarization phenomena in paramagnetic systems and will stimulate further work in this new direction.

In summary, we observe the EP in a hybrid quantum system consisting of P1 centers in diamond coupled to a coplanar-waveguide resonator and also convincingly show the cross-relaxation effect in P1 centers.
Among various applications of EPs, enhancing the sensitivity of detection has been widely investigated~\cite{PRL203901,PRL110802,nature548187,nature548192,Cao19,Wu20,Khurgin20}. In critical quantum metrology, it is important to precisely tune the coupling. For example, in Ref.~\cite{Garbe20}, a metrological protocol is designed to probe one physical parameter (e.g., the spin frequency) of the Rabi model by slowly sweeping the coupling from zero to some desired value close to the critical point. Thus, the good tunability of the coupling in this hybrid quantum system may facilitate the application of EPs in metrology.

The cross relaxation is an important phenomenon for nitrogen impurities in diamond, because it can be used to produce population inversion in paramagnetic nitrogen donors to implement solid-state masers~\cite{PhysRev.114.445,Sorokin60,Breeze18}.
In the future, one can introduce an effective gain in the coplanar-waveguide resonator~\cite{PhysRevA97} to balance the loss and gain in the hybrid system. This can achieve a hybrid system with parity-time symmetry. Also, in the present work, only stationary-state properties of the hybrid system are studied. Thus, another future work can be focused on the time evolution of the non-Hermitian system. While coalescent eigenvectors at the EP are guaranteed by the non-Hermiticity of the Hamiltonian~\cite{EP1}, it is interesting to probe the behavior of the eigenstates around the EP~\cite{hu,JDO}. For example, with the good tunability of the coupling strength and the magnon-mode frequency, it is promising to explore the exotic topological phenomena related to the EP (e.g., nonreciprocal energy transfer~\cite{hu} and asymmetric mode switching~\cite{JDO}) by dynamically encircling the EP via varying the coupling strength and the magnon-mode frequency in the hybrid system. Moreover, monitoring the evolution of the non-Hermitian system can also reveal intriguing time-dependent responses of the system on the drive tone, including the time evolution of the magnon occupation in each spin subensemble. This can provide further information about the cross relaxation in P1 centers.

\section*{Acknowledgments}

This work is supported by the National Key Research and Development Program of China (Grant No. 2016YFA0301200), the National Natural Science Foundation of China (Grants No. U1801661, No. 11934010, and No. 11774022), the Zhejiang Province Program for Science and Technology (Grant No. 2020C01019), the Science Challenge Project (Grant No. TZ2018003), the Beijing Academy of Quantum Information Sciences (BAQIS) Research Program (Grant No. Y18G27), and the China Postdoctoral Science Foundation (Grant No. 2020M671687). F.N. is supported in part by Nippon Telegraph and Telephone Corporation (NTT) Research, the Army Research Office (ARO) (Grant No. W911NF-18-1-0358), the Japan Science and Technology Agency (JST) [via the Quantum Leap Flagship Program (Q-LEAP) program and the Centers of Research Excellence in Science and Technology (CREST) Grant No. JPMJCR1676], the Japan Society for the Promotion of Science (JSPS) (via the Grants-in-Aid for Scientific Research (KAKENHI) Grant No. JP20H00134 and the Japan Society for the Promotion of Science (JSPS)-Russian Foundation for Basic Research (RFBR) Grant No. JPJSBP120194828), the Asian Office of Aerospace Research and Development (AOARD) (via Grant No. FA2386-20-1-4069), and the Foundational Questions Institute Fund (FQXi) via Grant No. FQXi-IAF19-06.

\vspace{.5cm}
\appendix

\section{EXPERIMENTAL SETUP}\label{appendix-A}
\begin{figure}[!hbt]
\centering
\includegraphics[scale=0.57]{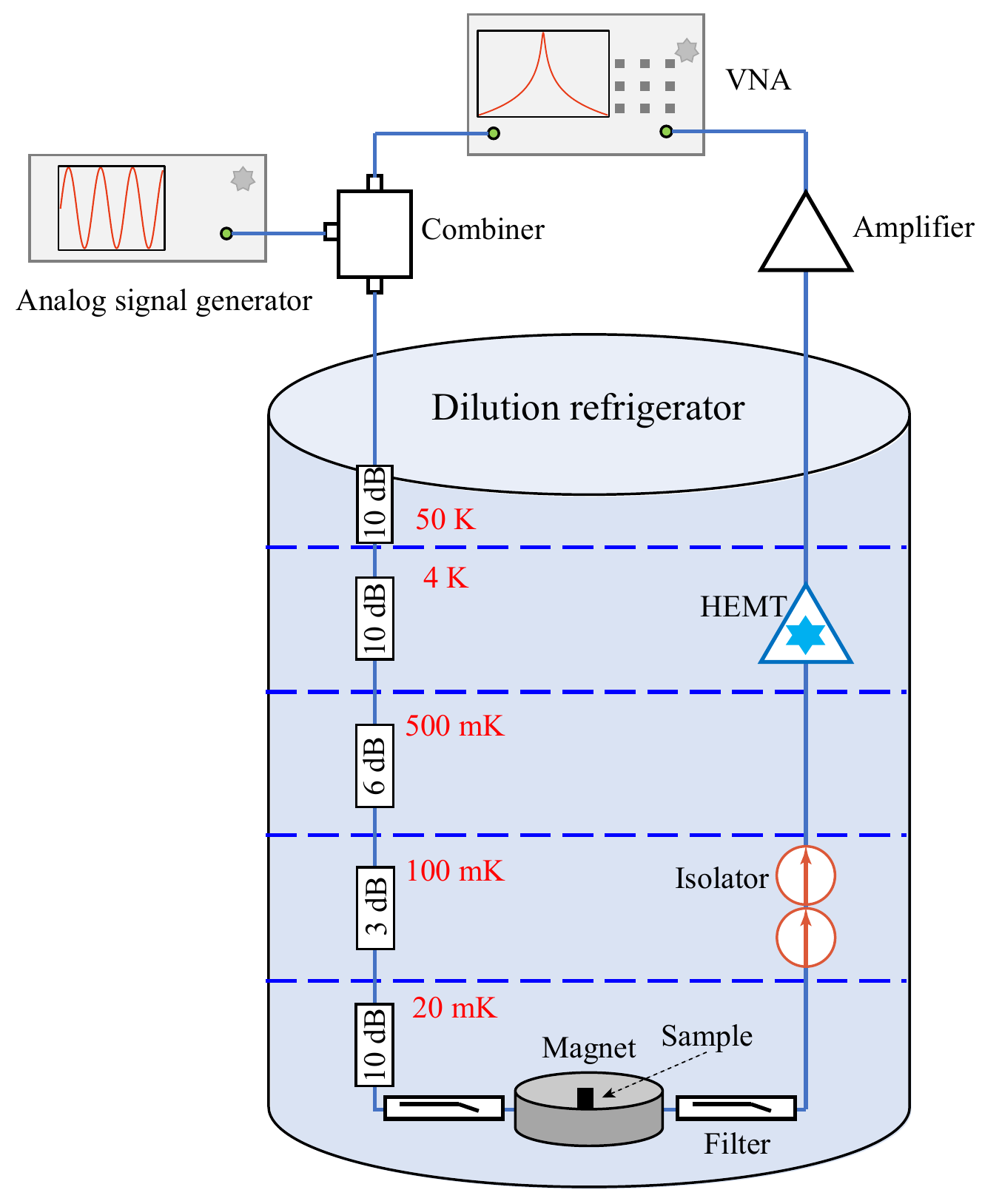}
\caption{A schematic of the experimental setup, where the hybrid system in Fig.~\ref{fig-sample}(a) is placed in a cryogenic chamber.}
\label{fig-setup}
\end{figure}

The hybrid system consists of an ensemble of P1 centers in diamond coupled to a coplanar waveguide resonator [Fig.~\ref{fig-sample}(a)]. The P1 center is a substitutional nitrogen defect [Fig.~\ref{fig-sample}(b)]. Due to the hyperfine interaction with the host $^14$N nucleus, the P1 center has six energy levels and three allowed transitions [Fig.~\ref{fig-sample}(c)]. In the experiment, the sample is placed in a dilution refrigerator and cooled down to 20 mK [Fig.~\ref{fig-setup}].

The coplanar waveguide resonator is fabricated by reactive ion etching of a 50-nm-thick d.c.-magnetron-sputtered niobium film on a thermally oxidized silicon substrate. The central conductor of the resonator is 20~$\mu$m wide and its gap to the ground plane is 11.6~$\mu$m, so that a 50~$\Omega$ characteristic impedance is obtained. The resonator has a length of 20 mm, defined by two nearly identical interdigital coupling capacitors with a capacitance of approximately 12 fF. The type-1b diamond used is synthesized under both high pressure and high temperature, in which the P1 centers are the main defects and constitute the spin ensembles harnessed in the experiment. The damping rate of the collective spin excitations (i.e., magnons) of the $s=0$ spin subensemble in diamond is about $\gamma/2\pi = 11.9\pm 0.3$ MHz (half width at half maximum). This is determined via the line width $\Gamma$ (half width at half maximum) of the polaritonic peaks under weak probe-field measurements [Fig.~\ref{fig-EP}(a)], with the relation $\Gamma = (\gamma+\kappa)/2$. The damping rates of the other two subensembles are nearly the same as that of the $s=0$ subensemble. To perform the measurement, we first apply a drive tone of a given frequency on the coplanar waveguide resonator for a duration of time and then apply a fast and low probe-power signal (approximately 1 fW) to implement the measurement of the transmission spectrum via a vector-network analyzer (VNA). The drive tone is generated by an analog signal generator and the duration time of the drive tone is set to be 3000~s to ensure that the system is in the stationary state. Then, we change the power of the drive tone and repeat the above process.

\section{EFFECTIVE HAMILTONIAN OF THE SYSTEM}\label{appendix-B}

The Hamiltonian of the P1 center includes the Zeeman energy and the hyperfine interaction between the electron (spin 1/2) and the host nucleus (spin 1)~\cite{loubser_electron_1978},
\begin{equation}\label{hamil-1}
H'_{\rm{P1}}=\gamma_{e}\mathbf{B} \cdot \mathbf{S}+\mathbf{S} \cdot A \cdot \mathbf{I},
\end{equation}
where $\gamma_{e}/2\pi=28$ GHz/T is the gyromagnetic ratio, $\mathbf{B}$ is the static magnetic field,
$\mathbf{S}\equiv(S_{x},S_{y},S_{z})$ and $\mathbf{I}\equiv(I_{x},I_{y},I_{z})$
are spin operators of the electron and the host nitrogen nucleus, respectively, and $A$ is the hyperfine interaction tensor~\cite{and_electron_1966}. Note that in Eq.~(\ref{hamil-1}), we leave out the Zeeman energy and the quadrupole interaction of the host nucleus, which only involve the operator $\mathbf{I}$~\cite{Lange12}. Due to the Jahn-Teller distortions, one of the four $\text{C}-\text{N}$ bonds is elongated. When the Jahn-Teller axis (i.e., the elongated $\text{C}-\text{N}$ bond) is along the $z$ direction, the corresponding hyperfine tensor $A$ is diagonal, i.e., $A_{\rm diag}/2\pi=\rm{diag}(81.8,81.8,114.2)$~MHz~\cite{and_electron_1966}.

In our experiment, the [100] crystal axis of the diamond sample is aligned along the external magnetic field $\mathbf{B}=B_{0}e_{z}$. Because all $\text{C}-\text{N}$ bonds have equivalent angles with $\mathbf{B}$, the hyperfine interaction is the same for each of the P1 centers. In the coordinate frame with the $z$ axis oriented along the [100] crystal axis, the hyperfine tensor $A$ becomes off diagonal, which can be obtained via a transformation of the diagonal hyperfine tensor $A_{\rm diag}$~\cite{Degen20}.
The Hamiltonian of the P1 center can be rewritten as
\begin{equation}\label{hamil-2}
H'_{\rm{P1}}=\gamma_{e}B_{0}S_{z}+A_{\parallel}S_{z}I_{z}+A_{\perp}(S_{x}I_{x}+S_{y}I_{y})+\mathbf{S} \cdot A_{\rm off} \cdot \mathbf{I},
\end{equation}
where $A_{\parallel}/2\pi = A_{\perp}/2\pi \approx 94$~MHz are the diagonal elements of the hyperfine tensor $A$ and the off-diagonal elements are included in the tensor $A_{\rm off}$. The hyperfine interaction is much smaller than the Zeeman energy and can be treated as a perturbation. From first-order perturbation theory, the Hamiltonian is reduced to
\begin{equation}
H_{\rm{P1}} = \langle s | H'_{\rm{P1}} | s \rangle=(\gamma_{e}B_{0}+A_{\parallel}s)S_{z},
\end{equation}
where $s=0$, $\pm 1$ are the three eigenvalues of the operator $I_{z}$, with $|s\rangle$ being the corresponding eigenvectors. According to the values of $s$, the ensemble of P1 centers can be divided into three subensembles of spins with transition frequencies $\omega_{+}=\gamma_{e}B_{0}+A_{\parallel}$, $\omega_{0}=\gamma_{e}B_{0}$, and $\omega_{-}=\gamma_{e}B_{0}-A_{\parallel}$,  respectively.

If we only take one subensemble into consideration (e.g., $s=0$), the Hamiltonian of the hybrid system is
\begin{equation}\label{hamil-3}
\begin{aligned}
H_{s}=\omega_{c}a^{\dag}a+\sum_{i=1}^N\omega_{0}S_{z}^{i}
+\sum_{i=1}^Ng_{s,i}(aS_{i}^{+}+a^{\dag}S_{i}^{-}),
     \end{aligned}
\end{equation}
where $a$ ($a^{\dag}$) is the annihilation (creation) operator of the resonator mode with frequency $\omega_{c}$, $\mathbf{S}_{i}\equiv(S_{x}^{i},S_{y}^{i},S_{z}^{i})$, and $S_{i}^{\pm}\equiv S_{x}^{i} \pm iS_{y}^{i}$ are the operators of the $i$th spin in the $s=0$ subensemble, and $g_{s,i}$ is the coupling strength between the $i$th spin and the resonator mode. For simplicity, the same coupling strength $g_{s,i} = g_s$ is assumed for all spins in the subensemble. To describe the collective behavior of the spins, we define the macrospin operator $\mathbf{J}\equiv (J_{x},J_{y},J_{z})=\sum_{i}\mathbf{S}_{i}$ and the collective coupling strength $g\equiv \sqrt{N}g_s$, with $N$ being the number of spins in the subensemble. Then, the Hamiltonian (\ref{hamil-3}) is reduced to
\begin{equation}
\label{TC}
H_{s}=\omega_{c}a^{\dag}a+\omega_{0}J_{z}
+\frac{g}{\sqrt{N}}(aJ_{+}+a^{\dag}J_{-}),
\end{equation}
with $J_{\pm}\equiv J_x\pm iJ_y$.

To study the exceptional point (EP) of the hybrid system, we first use the Holstein-Primakoff transformation~\cite{Holstein40},
\begin{eqnarray}\label{}
J_{+}&=&b^{\dagger}\sqrt{N-b^{\dagger}b},\nonumber\\
J_{-}&=&\sqrt{N-b^{\dagger}b}\,b,\\
J_{z}&=&b^{\dagger}b-N/2,\nonumber
\end{eqnarray}
to convert the Hamiltonian (\ref{TC}) to
\begin{eqnarray}\label{bos}
H_{s}&=&\omega_{c}a^{\dag}a+\omega_{0}b^{\dagger}b \\
     & &+g\left(b^{\dagger}\sqrt{1-b^{\dagger}b/N}\,a+a^{\dagger}\sqrt{1-b^{\dagger}b/N}\,b\right),\nonumber
\end{eqnarray}
where $b$ ($b^{\dagger}$) is the annihilation (creation) operator of the magnons, which are the collective spin excitations in the $s=0$ subensemble. Next, we linearize the above Hamiltonian under the mean-field approximation. Based on the Taylor's expansion in terms of $b^{\dagger}b/N$, the coupling term $a^{\dagger}\sqrt{1-b^{\dagger}b/N}\,b$ can be written as
\begin{eqnarray}\label{expand}
& &a^{\dagger}\sqrt{1-b^{\dagger}b/N}\,b\nonumber\\
& &=a^{\dagger}[1-b^{\dagger}b/(2N)-(b^{\dagger}b)^{2}/(8N^{2})
                                       -(b^{\dagger}b)^{3}/(16N^{3})+\cdot\cdot\cdot \nonumber\\
& &~~~~        +(b^{\dagger}b/N)^{n}f^{(n)}/n!+\cdot\cdot\cdot]b\nonumber\\
& &=a^{\dagger}b-a^{\dagger}b^{\dagger}bb/(2N)-(b^{\dagger}b)a^{\dagger}b^{\dagger}bb/(8N^{2})\nonumber\\
& &~~~~      -(b^{\dagger}b)^2a^{\dagger}b^{\dagger}bb/(16N^{3})+\cdot\cdot\cdot \nonumber\\
& &~~~~      +(b^{\dagger}b)^{n-1}a^{\dagger}b^{\dagger}bb(1/N^{n})f^{(n)}/n!+\cdot\cdot\cdot
\end{eqnarray}
where $f^{(n)}=-(2n-3)!!/2^{n}$, with $n \geq 2$. We write $b=\beta+\delta b$ and $b^{\dag}=\beta^{*}+\delta b^{\dag}$, where $\beta$ ($\beta^{*}$) is the mean value of the operator $b$ ($b^{\dag}$) and $\delta b$ ($\delta b^{\dag}$) is the corresponding fluctuation. When keeping the terms up to first-order fluctuations,
\begin{eqnarray}\label{1st}
a^{\dagger}b^{\dagger}bb&=&a^{\dagger}(\beta^{*}+\delta b^{\dag})(\beta+\delta b)(\beta+\delta b) \nonumber\\
&\approx& a^{\dagger}(2|\beta|^{2} b+\beta^{2} b^{\dag}-2|\beta|^{2}\beta).
\end{eqnarray}
Neglecting the counter-rotating term $a^{\dagger}b^{\dag}$ under the rotating-wave approximation, we have
\begin{equation}\label{1st-1}
a^{\dagger}b^{\dagger}bb\approx 2|\beta|^{2}a^{\dagger}b-2|\beta|^{2}\beta a^{\dagger}.
\end{equation}
Also, we have
\begin{eqnarray}\label{2nd}
(b^{\dagger}b)a^{\dagger}b^{\dagger}bb
&\approx&b^{\dagger}b(2|\beta|^{2} a^{\dagger}b-2|\beta|^{2}\beta a^{\dagger}) \nonumber\\
&=&2|\beta|^{2} a^{\dagger}b^{\dagger}bb-2|\beta|^{2}\beta b^{\dagger}ba^{\dagger} \nonumber\\
&\approx&2|\beta|^{2}(2|\beta|^{2} a^{\dagger}b-2|\beta|^{2}\beta a^{\dagger}) -2|\beta|^{4}\beta a^{\dagger} \nonumber\\
&=&(2|\beta|^{2})^2a^{\dagger}b-6|\beta|^{4}\beta a^{\dagger},
\end{eqnarray}
\begin{eqnarray}\label{3rd}
(b^{\dagger}b)^2a^{\dagger}b^{\dagger}bb
&=&(b^{\dagger}b)(b^{\dagger}b)a^{\dagger}b^{\dagger}bb \nonumber\\
&\approx&(b^{\dagger}b)\left[(2|\beta|^{2})^2a^{\dagger}b-6|\beta|^{4}\beta a^{\dagger}\right] \nonumber\\
&=&(2|\beta|^{2})^2 a^{\dagger}b^{\dagger}bb-6|\beta|^{4}\beta b^{\dagger}ba^{\dagger} \nonumber\\
&\approx&(2|\beta|^{2})^2(2|\beta|^{2} a^{\dagger}b-2|\beta|^{2}\beta a^{\dagger}) -6|\beta|^{6}\beta a^{\dagger} \nonumber\\
&=&(2|\beta|^{2})^3a^{\dagger}b-14|\beta|^{6}\beta a^{\dagger}.
\end{eqnarray}
More generally, we have
\begin{eqnarray}\label{nth}
(b^{\dagger}b)^{n-1}a^{\dagger}b^{\dagger}bb
\approx(2|\beta|^{2})^{n}a^{\dagger}b-\sum_{k=1}^{n}2^{k}|\beta|^{2n}\beta a^{\dagger}.~~~
\end{eqnarray}
Therefore, Eq.~(\ref{expand}) can be approximately written as
\begin{eqnarray}\label{expand-1}
&&a^{\dagger}\sqrt{1-b^{\dagger}b/N}\,b\nonumber\\
&&=\big[1-2|\beta|^{2}/(2N)-(2|\beta|^{2})^2/(8N^{2})-(2|\beta|^{2})^3/(16N^{3})\nonumber\\
&&~~~~+\cdot\cdot\cdot+(2|\beta|^{2}/N)^{n}f^{(n)}/n!+\cdot\cdot\cdot\big]a^{\dagger}b+(\Omega_{b}/g)a^{\dagger} \nonumber\\
&&=\sqrt{1-2|\beta|^{2}/N}\,a^{\dagger}b+(\Omega_{b}/g)a^{\dagger},
\end{eqnarray}
where
\begin{eqnarray}\label{biased}
\Omega_{b}/g&=&2|\beta|^{2}\beta/(2N)+6|\beta|^{4}\beta/(8N^{2})+14|\beta|^{6}\beta/(16N^{3})\nonumber\\
            & &+\cdot\cdot\cdot-f^{(n)}/n!\sum_{k=1}^{n}2^{k}|\beta|^{2n}\beta/N^{n}+\cdot\cdot\cdot.
\end{eqnarray}
When including the coefficient of the counter-rotating terms, the neglected counter-rotating part in the second term of the Taylor's expansion [cf. Eqs.~(\ref{bos})-(\ref{1st})] is $c_{1}ga^{\dag}b^{\dag}$, where $c_{1}=-\beta^{2}/2N$. For the third term of the Taylor's expansion, the neglected counter-rotating part is $c_{2}ga^{\dag}b^{\dag}$, where $c_{2}=-|\beta|^{2}\beta^{2}/8N^{2}$. In fact, the neglected counter-rotating parts in all terms of the Taylor's expansion can be summed as $gC_{\Sigma}a^{\dag}b^{\dag}+{\rm H.c.}$, where
\begin{eqnarray}\label{}
C_{\Sigma}&=&-\beta^{2}/2N-|\beta|^{2}\beta^{2}/8N^{2}-|\beta|^{4}\beta^{2}/16N^{3}+\cdot\cdot\cdot\nonumber\\
           & &+(|\beta|^{2(n-1)}\beta^{2}/N^{n})f^{(n)}/n!+\cdot\cdot\cdot\nonumber\\
           &=&-\Big(1-\sqrt{1-|\beta|^{2}/N}\Big)\beta^{2}/|\beta|^{2}.
\end{eqnarray}
Obviously, $|C_{\Sigma}|\equiv 1-\sqrt{1-|\beta|^{2}/N}<1$. Because $g\ll\omega_c,\omega_0$ in our hybrid system, it is thus reasonable to perform the rotating-wave approximation in the above derivations, even under high pump powers. Since all high-order terms in the Taylor's expansion are considered in Eqs.~(\ref{expand})-(\ref{biased}), the mean-field approximation in Eq.~(\ref{expand-1}) is also valid for the high-excitation case.

Substituting Eq.~(\ref{expand-1}) and its Hermitian conjugate into Eq.~(\ref{bos}), we obtain
\begin{equation}\label{sys}
H_{s}=\omega_{c}a^{\dag}a+\omega_{0}b^{\dagger}b
      +g_{\rm eff}(a^{\dagger}b+ab^{\dagger})+(\Omega_{b}a^{\dagger}+\Omega^{*}_{b}a),
\end{equation}
where $g_{\rm eff}=g\sqrt{1-|\beta|^{2}/(N/2)}$. Obviously, the last term in the above equation can be absorbed into the drive Hamiltonian $H_{d}$:
\begin{equation}\label{}
H_{d}=\Omega_{d}a^{\dag}e^{-i\omega_{d}t}+\Omega^{*}_{d}ae^{i\omega_{d}t}.
\end{equation}
Actually, the displacement term $\Omega_{b}a^{\dagger}+\Omega^{*}_{b}a$ in the Hamiltonian (\ref{sys}) is a counter-rotating term. When the magnons are tuned to be nearly resonant with the resonator mode (i.e., $\omega_0\sim\omega_c$), this counter-rotating term can also be ignored and the Hamiltonian (\ref{sys}) is reduced to $H_{s}=\omega_{c}a^{\dag}a+\omega_{0}b^{\dagger}b+g_{\rm eff}(a^{\dagger}b+ab^{\dagger})$ in the near-resonance case.
When the decay rates of the resonator mode and the spin ensemble, $\kappa$ and $\gamma$, are included, the Hamiltonian of the system can be effectively written, in the non-Hermitian form~\cite{Nori19,Nori20}, as
\begin{equation}\label{tot}
H_{\rm eff}=(\omega_{c}-i\kappa)a^{\dag}a+(\omega_{0}-i\gamma)b^{\dagger}b+g_{\rm eff}(a^{\dagger}b+ab^{\dagger}).
\end{equation}
With the obtained effective non-Hermitian Hamiltonian in Eq.~(\ref{tot}), the dynamics of the hybrid system is governed by the following Liouvillian equation~\cite{Nori19}:
\begin{equation}
\frac{\partial \rho(t)}{\partial t}=-i[H_{\rm eff}\rho(t)-\rho(t)H_{\rm eff}^{\dag}],
\end{equation}
where $\rho(t)$ is the density operator of the system at time $t$. Solving the Liouvillian equation, we can express the density operator as
\begin{equation}
\rho(t)=\exp(-iH_{\rm eff}t)\rho(0)\exp(iH_{\rm eff}^{\dag}t).
\end{equation}
It is known in quantum mechanics that $\rho^2(t)=\rho(t)$ if the system is in a pure state. However, in the non-Hermitian case that we study,
\begin{eqnarray}
\rho^2(t)&=&\exp(-iH_{\rm eff}t)\rho(0)\exp(iH_{\rm eff}^{\dag}t)\exp(-iH_{\rm eff}t)\nonumber\\
&&\times \rho(0)\exp(iH_{\rm eff}^{\dag}t).
\end{eqnarray}
Due to $H_{\rm eff}^{\dag}\neq H_{\rm eff}$, $\exp(iH_{\rm eff}^{\dag}t)\exp(-iH_{\rm eff}t)\neq 1$, so there is $\rho^2(t)\neq\rho(t)$, even if the non-Hermitian Hamiltonian has a simple form in Eq.~(\ref{tot}). This means that the system is in a mixed state, instead of a pure state.

\begin{figure}[!h]
\includegraphics[width=0.44\textwidth]{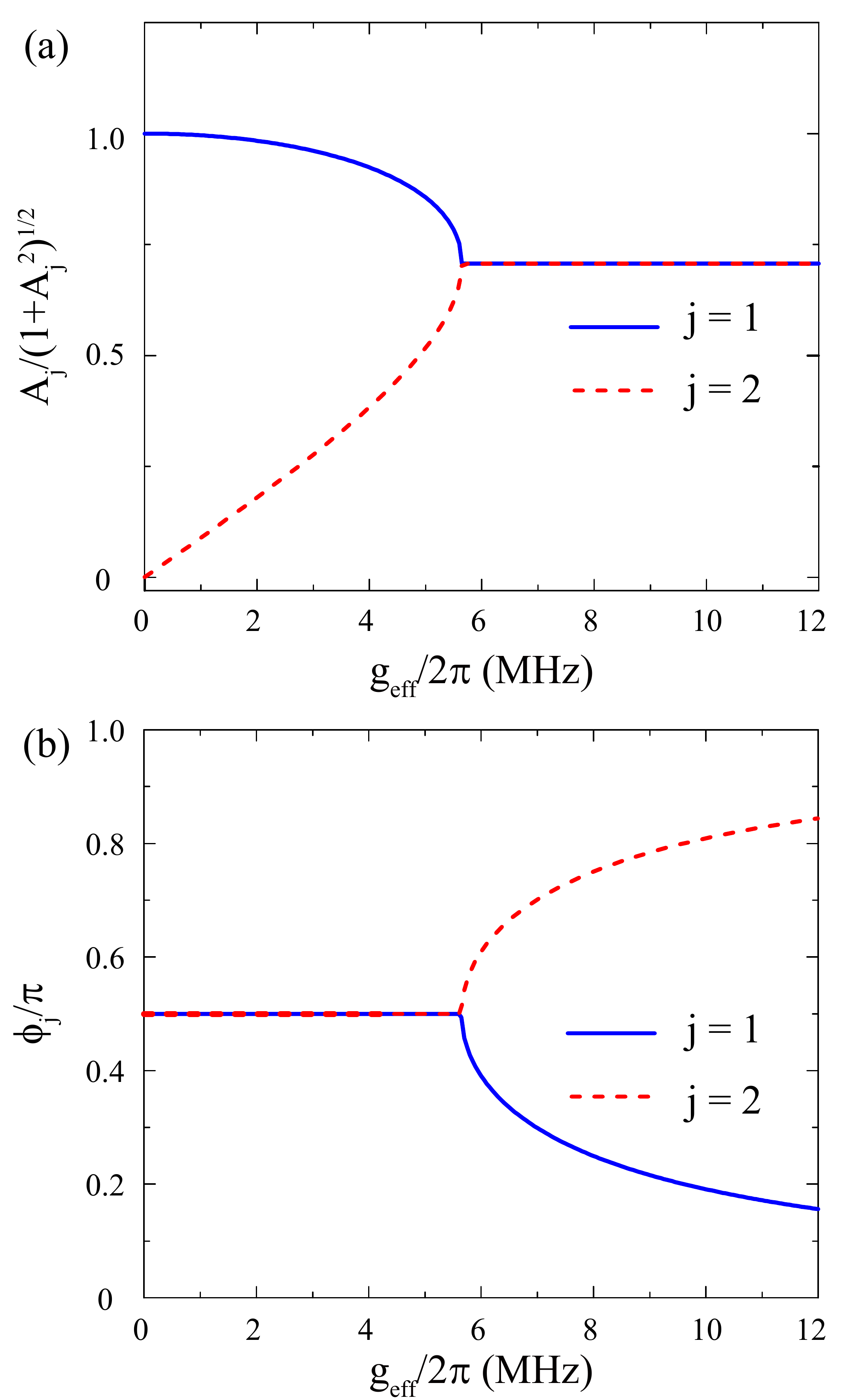}
\caption{The effective-coupling dependence of the parameters of the eigenvector. (a) The normalized amplitudes $A_{1,2}/\sqrt{1+A_{1,2}^2}$ and (b) the relative phases $\phi_{1,2}/\pi$ of the two eigenvectors versus the effective coupling strength $g_{\rm eff}/2\pi$. Here, we choose $\kappa/2\pi=0.6$~MHz and $\gamma/2\pi=11.9$~MHz, as in the main text.}
\label{fig-eigen}
\end{figure}

In matrix form, we can write the effective non-Hermitian Hamiltonian as
\begin{equation}
\label{matrix-1}
H_{\rm eff}=\begin{pmatrix}
  \omega_{c}-i\kappa & g_{\rm eff}  \\
  g_{\rm eff} & \omega_{0}-i\gamma  \\
 \end{pmatrix}.
\end{equation}
Diagonalizing it, we obtain the two eigenvalues,
\begin{eqnarray}\label{eigen}
\omega_{1,2}&=&\frac{1}{2}[(\omega_{c}+\omega_{0})-i(\kappa+\gamma)]\nonumber\\
            & &\pm\frac{1}{2}\sqrt{4g_{\rm eff}^{2}+[(\omega_{c}-i\kappa)-(\omega_{0}-i\gamma)]^{2}},
\end{eqnarray}
this being Eq.~(\ref{eigenvalues}) in the main text. In the resonant case ($\omega_c=\omega_0$) that we study, Eq.~(\ref{eigen}) reduces to Eq.~(\ref{EQ3}) in the main text, i.e.,
\begin{equation}
\label{eigen1}
\omega_{1,2}=\omega_{0}-\frac{1}{2}i(\kappa+\gamma)\pm\frac{1}{2}\sqrt{4g_{\rm eff}^2-(\gamma-\kappa)^2}.
\end{equation}
Corresponding to these two eigenvalues, the two eigenvectors are
\begin{eqnarray}
\label{eigenvectors}
|\psi_{1,2}\rangle&=&\Bigg(\frac{i(\gamma-\kappa)\pm\sqrt{4g_{\rm eff}^2-(\gamma-\kappa)^2}}{2g_{\rm eff}},1\Bigg)^{T}\nonumber\\
                &\equiv& \big(A_{1,2}e^{i\phi_{1,2}},1\big)^{T},
\end{eqnarray}
where $T$ denotes the matrix transpose and $\gamma>\kappa$ in our hybrid system. The relative amplitudes $A_{1,2}$ can be written as
\begin{equation}\label{lambda}
\begin{split}
A_{1,2}
=\Bigg\{
\begin{array}{cc}
                \frac{(\gamma-\kappa)\pm
                   \sqrt{(\gamma-\kappa)^2-4g_{\rm eff}^2}}{2g_{\rm eff}},
                                                      &~~~g_{\rm eff}\leq(\gamma-\kappa)/2;\\
                         1,          &~~~g_{\rm eff}>(\gamma-\kappa)/2,
\end{array}
\end{split}
\end{equation}
and the relative phases $\phi_{1,2}$ are
\begin{equation}\label{phi}
\begin{split}
\phi_{1,2}
=\Bigg\{
\begin{array}{cc}
                 \pi/2,                         & g_{\rm eff}\leq(\gamma-\kappa)/2;\\
  \arccos\left[\pm\frac{\sqrt{4g_{\rm eff}^2-(\gamma-\kappa)^2}}{2g_{\rm eff}}\right],  & g_{\rm eff}>(\gamma-\kappa)/2.
\end{array}
\end{split}
\end{equation}
Obviously, the two eigenvectors coalesce to $|\psi_{1}\rangle=|\psi_{2}\rangle=(i,1)^T$ at the EP:~$g_{\rm eff}=(\gamma-\kappa)/2$.
On the left side of the EP, i.e., $g_{\rm eff}<(\gamma-\kappa)/2$, the relative amplitudes $A_{1,2}$ of the two eigenvectors are different, but the relative phases $\phi_{1,2}$ are the same. On the contrary, on the right side of the EP, i.e., $g_{\rm eff}>(\gamma-\kappa)/2$, the relative amplitudes $A_{1,2}$ are the same, but the relative phases $\phi_{1,2}$ are different (cf. Fig.~\ref{fig-eigen}).

\section{STEADY-STATE MEASUREMENT AND DEPENDENCE OF THE RESONATOR-MODE FREQUENCY ON THE STATIC MAGNETIC FIELD}\label{appendix-C}

\begin{figure}[!hbt]
\centering
\includegraphics[scale=0.37]{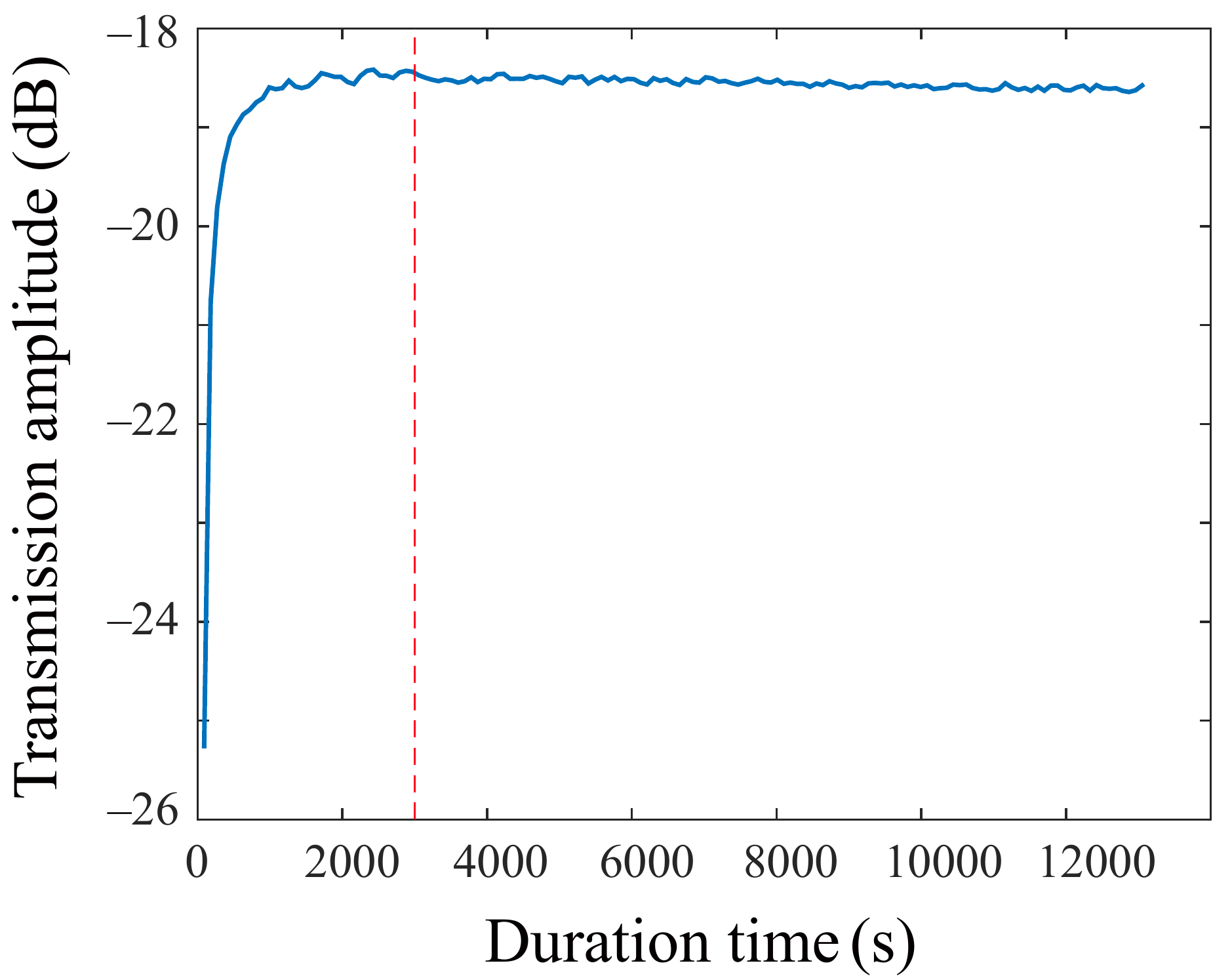}
\caption{The transmission amplitude versus the duration time of the drive tone. The $s=0$ subensemble is tuned to be in resonance with the resonator mode. We resonantly pump the $s=0$ subensemble with a drive tone and measure the transmission amplitude at 3.090~GHz. To measure the transmission spectra in our experiment, the duration time of the drive tone is chosen to be 3000~s before each measurement of the transmission amplitude is implemented.}
\label{fig-amplitude}
\end{figure}

To measure the transmission spectra in Figs.~\ref{fig-EP}(b), \ref{fig-cross}(a), and \ref{fig-cross}(b) (see the main text), we first tune the $s=0$ subensemble in resonance with the resonator mode $\omega_{0}/2\pi=\omega_{c}/2\pi= 3.093$~GHz and then pump the system with a drive tone at $\omega_{d}/2\pi= 3.093$~GHz for a duration of time. Meanwhile, we monitor the transmission amplitude at 3.090 GHz and measure it once every 5 seconds. As shown in Fig.~\ref{fig-amplitude}, the transmission amplitude becomes nearly stable when the duration time of the drive tone is increased to be approximately $2000$~s. To ensure that the hybrid system is in the steady state, in both Fig.~\ref{fig-EP}(b) and Fig.~\ref{fig-cross} the duration time of the drive tone is chosen to be 3000s before each measurement of the transmission amplitude is implemented (cf. the dashed vertical line in Fig.~\ref{fig-amplitude}).

\begin{figure*}[!hbt]
\centering
\includegraphics[scale=0.37]{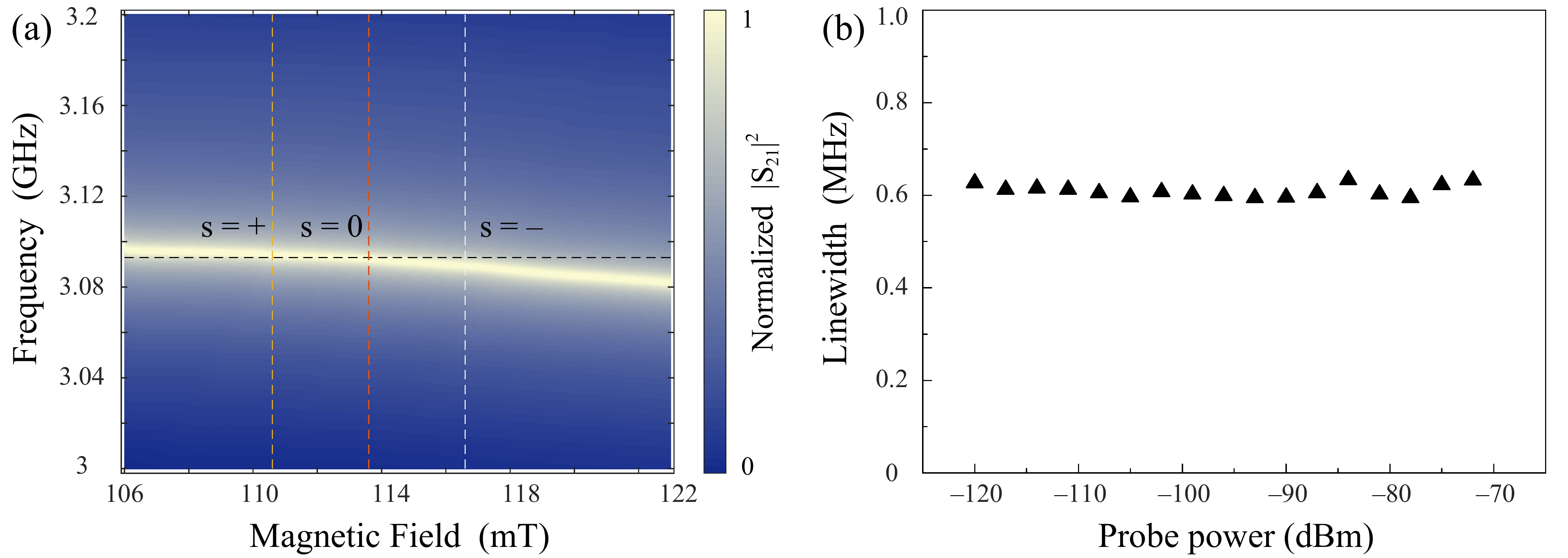}
\caption{(a) The dependence of the resonator-mode bare frequency on the static magnetic field. The transmission spectrum is obtained in the same way as in Fig.~\ref{fig-EP}(a), but using a strong probe tone. (b) The measured line width of the resonator mode versus the drive power.}
\label{fig-frequency}
\end{figure*}

In the cases of Figs.~\ref{fig-cross}(c) and \ref{fig-cross}(d), the bare frequencies of the resonator mode are fitted to be $3.095$~GHz and $3.090$~GHz, respectively. The different bare resonator-mode frequencies found in Figs.~\ref{fig-cross}(c) and \ref{fig-cross}(d) are due to the effect of the static magnetic field on the superconducting waveguide resonator, because the applied static magnetic field can unavoidably affect the superconductivity of the resonator. To verify this, we measure the dependence of the transmission spectrum on the magnetic field in the same way as in Fig.~\ref{fig-EP}(a), but the probe tone is chosen to be so intense that all three spin subensembles are almost decoupled from the resonator, i.e. the magnon occupation is achieved to be $\langle b^{\dag}b\rangle=(N/2)$ for each subensemble to have $g_{\rm eff}=g_{\rm eff,\pm}=0$. Indeed, as shown in Fig.~\ref{fig-frequency}(a), the bare frequency of the resonator mode is gradually shifted as the static magnetic field strengthens. The three dashed vertical lines (from the left to the right) correspond to the magnetic-field strengths used in Fig.~\ref{fig-cross}(c), Fig.~\ref{fig-EP}(b) [i.e., Figs.~\ref{fig-cross}(a) and \ref{fig-cross}(b)], and Fig.~\ref{fig-cross}(d), respectively. The black dashed horizontal line indicates the bare frequency of the resonator mode in the cases of Fig.~\ref{fig-EP}(b) and Figs.~\ref{fig-cross}(a) and \ref{fig-cross}(b).

In analyzing the experimental results, we assume that the decay rate of the resonator mode is independent of the drive power. To demonstrate this, in Fig.~\ref{fig-frequency}(b), we measure the decay rate of the resonator mode versus the drive power. It can be seen that the measured decay rate of the resonator fluctuates slightly around $\kappa/2\pi=0.6$~MHz. This indicates that the decay rate of the resonator is nearly independent of the drive power.

\section{STEADY-STATE SOLUTION OF THE MAGNON OCCUPATION}\label{appendix-D}

In the experiment, we apply a driving tone to achieve considerable magnon occupations in the considered spin subensemble. With the drive field included, the Hamiltonian of the system becomes
\begin{equation}\label{Hamil}
H=\Delta_{c}a^{\dag}a+\Delta_{s}J_{z}+\frac{g}{\sqrt{N}}(aJ_{+}+a^{\dag}J_{-})+(\Omega_{d}a^{\dag}+\Omega^{*}_{d}a),
\end{equation}
in the rotating frame with respect to the frequency $\omega_{d}$ of the drive field, where $\Delta_{c}=\omega_{c}-\omega_{d}$ ($\Delta_{s}=\omega_{s}-\omega_{d}$) is the frequency detuning between the resonator mode (spin subensemble) and the drive field. When the decay rates $\kappa$ and $\gamma$ of the resonator mode and the spin subensemble are considered, the dynamics of the hybrid system can be described using the quantum Langevin equations~\cite{Walls94}:
\begin{equation}\label{Langevin}
\begin{split}
&\dot{a}=-i(\Delta_{c}-i\kappa)a-i\frac{g}{\sqrt{N}}J_{-}-i\Omega_d,\\
&\dot{J_{-}}=-i(\Delta_{s}-i\gamma)J_{-}+i\frac{2g}{\sqrt{N}}J_{z}a.
\end{split}
\end{equation}
From Eq.~(\ref{Langevin}), it follows that the mean values of $a$ and $J_{-}$ satisfy
\begin{equation}\label{}
\begin{split}
&\langle\dot{a}\rangle=-i(\Delta_{c}-i\kappa)\langle a\rangle-i\frac{g}{\sqrt{N}}\langle J_{-}\rangle-i\Omega_d,\\
&\langle\dot{J_{-}}\rangle=-i(\Delta_{s}-i\gamma)\langle J_{-}\rangle+i\frac{2g}{\sqrt{N}}\langle J_{z}\rangle\langle a\rangle,
\end{split}
\end{equation}
where the mean-field approximation $\langle J_{z}a\rangle=\langle J_{z}\rangle\langle a\rangle$ is used for the two-operator term.

For the steady state of the system, $\langle {\dot a } \rangle =\langle {\dot J_{-} } \rangle =0$, so
\begin{equation}\label{steady}
\begin{split}
&(\Delta_{c}-i\kappa)\langle a\rangle+\frac{g}{\sqrt{N}}\langle J_{-}\rangle+\Omega_d=0,\\
&(\Delta_{s}-i\gamma)\langle J_{-}\rangle-\frac{2g}{\sqrt{N}}\langle J_{z}\rangle\langle a\rangle=0.
\end{split}
\end{equation}
The first equation in Eq.~(\ref{steady}) gives
\begin{equation}
\langle a\rangle=-\frac{g\langle J_{-}\rangle}{\sqrt{N}(\Delta_{c}-i\kappa)}-\frac{\Omega_d}{\Delta_{c}-i\kappa}.
\end{equation}
Substituting the above equation into the second equation in Eq.~(\ref{steady}), we have
\begin{equation}\label{solution-1}
(\Delta_{s}-i\gamma)\langle J_{-}\rangle+\frac{2g^2\langle J_{-}\rangle\langle J_{z}\rangle}{N(\Delta_{c}-i\kappa)}+\frac{2g\Omega_d\langle J_{z}\rangle}{\sqrt{N}(\Delta_{c}-i\kappa)}=0.
\end{equation}
With the Holstein-Primakoff transformation and the mean-field approximation, $\langle J_{z}\rangle$ and $\langle J_{-}\rangle$ can be expressed as
$\langle J_{z}\rangle=\langle b^{\dag}b\rangle-N/2$, and $\langle J_{-}\rangle=\langle\sqrt{(N- b^{\dag}b)b^2}\rangle\approx\sqrt{(N-\langle b^{\dag}b\rangle)\langle b^2\rangle}$.
Then, Eq.~(\ref{solution-1}) is converted to
\begin{equation}
\Big[(\Delta_{s}-\eta\Delta_{c})-i(\gamma+\eta\kappa)\Big]\sqrt{\langle b^2\rangle/N}
-\frac{g\sqrt{1-\chi}}{\Delta_{c}-i\kappa}\xi(\Omega_{d}/\sqrt{N})=0,
\end{equation}
where $\xi=(1-2\chi)/(1-\chi)$ and $\eta=g^2(1-2\chi)/(\Delta_{c}^2+\kappa^2)$, with $\chi=\langle b^{\dag}b\rangle/N$ being the reduced magnon occupation.
Since $\sqrt{\langle b^2\rangle/N}\;(\!\sqrt{\langle b^2\rangle/N})^*\approx \chi$, multiplying the above equation with its complex conjugate, we obtain the steady-state solution for the magnon occupation:
\begin{equation}\label{solution}
\left[(\Delta_{s}-\eta\Delta_{c})^{2}+(\gamma+\eta\kappa)^{2}\right]\chi-\xi\eta(|\Omega_{d}|/\sqrt{N})^2=0.
\end{equation}
To measure the transmission spectrum in Fig.~\ref{fig-EP}(b) of the main text, we use $\omega_{d}=\omega_{0}=\omega_{c}$, i.e. $\Delta_{s}=\Delta_{c}=0$. In this resonant case, Eq.~(\ref{solution}) is reduced to Eq.~(\ref{driving}) in the main text for fitting the effective magnon-photon coupling in Fig.~\ref{fig-EP}(c), where $\eta=g^2(1-2\chi)/\kappa^2=g^2_{\rm eff}/\kappa^2$.

\begin{figure}[!hbt]
\centering
\includegraphics[scale=0.4]{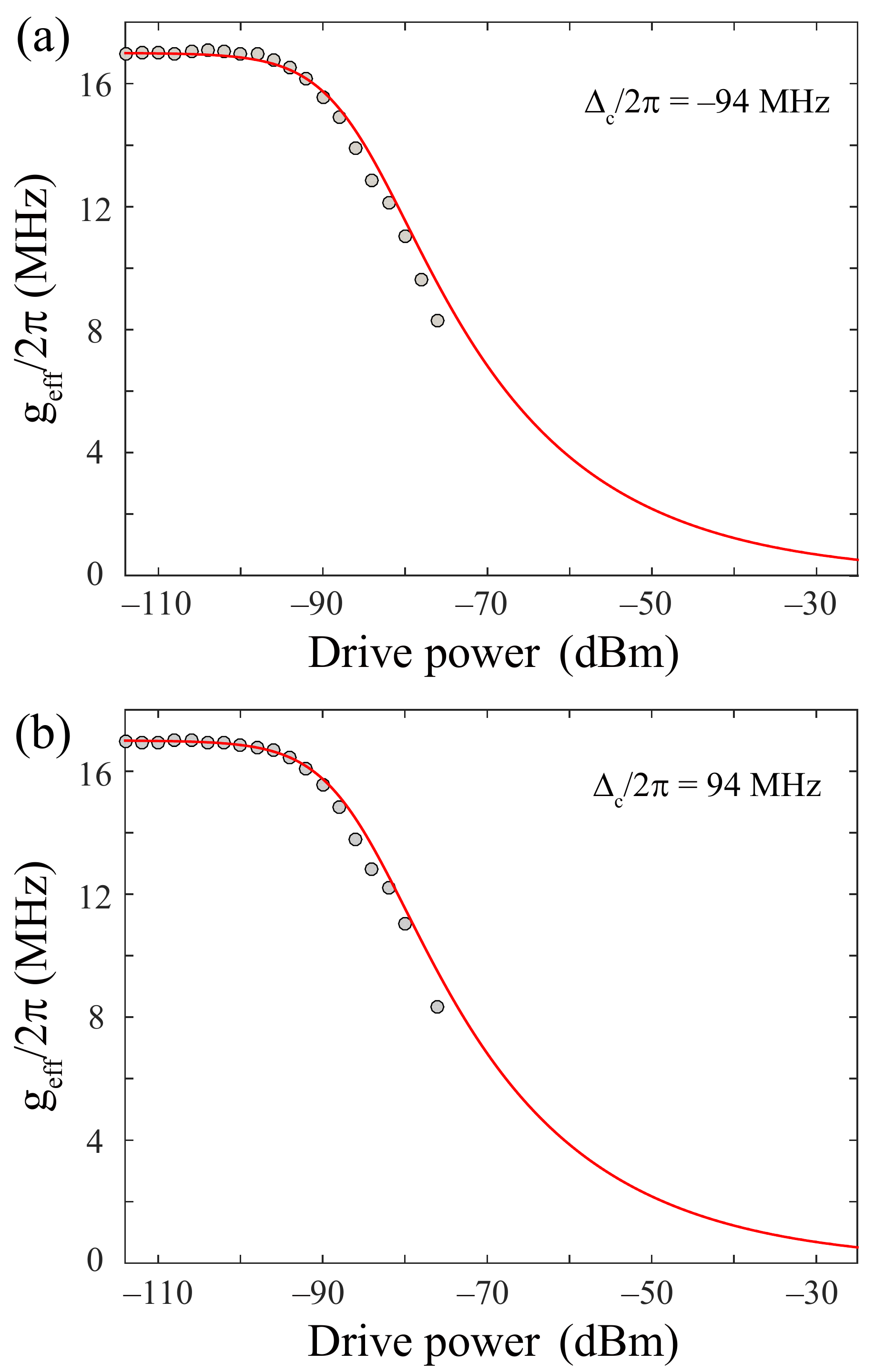}
\caption{The effective magnon-photon coupling versus the driving power. (a)~The data for $g_{\rm eff}$ (circles) extracted from the Rabi splitting in Fig.~\ref{fig-cross}(a) (see the main text), where the $s=0$ subensemble is in resonance with the resonator mode and the $s=+$ subensemble is resonantly pumped by a drive tone. The solid curve is calculated for $g_{\rm eff,+}$ using Eq.~(\ref{solution2}) with $\Delta_{c}/2\pi=-94$ MHz. (b)~The data for $g_{\rm eff}$ (circles) extracted from the Rabi splitting in Fig.~\ref{fig-cross}(b), where the $s=0$ subensemble is in resonance with the resonator mode and the $s=-$ subensemble is resonantly pumped by a drive tone. The solid curve is calculated for $g_{\rm eff,-}$ using Eq.~(\ref{solution2}) with $\Delta_{c}/2\pi=94$ MHz. The other parameters are the same as in Fig.~\ref{fig-EP}(c).}
\label{fig-coupling}
\end{figure}

\begin{figure*}[!hbt]
\centering
\includegraphics[scale=0.4]{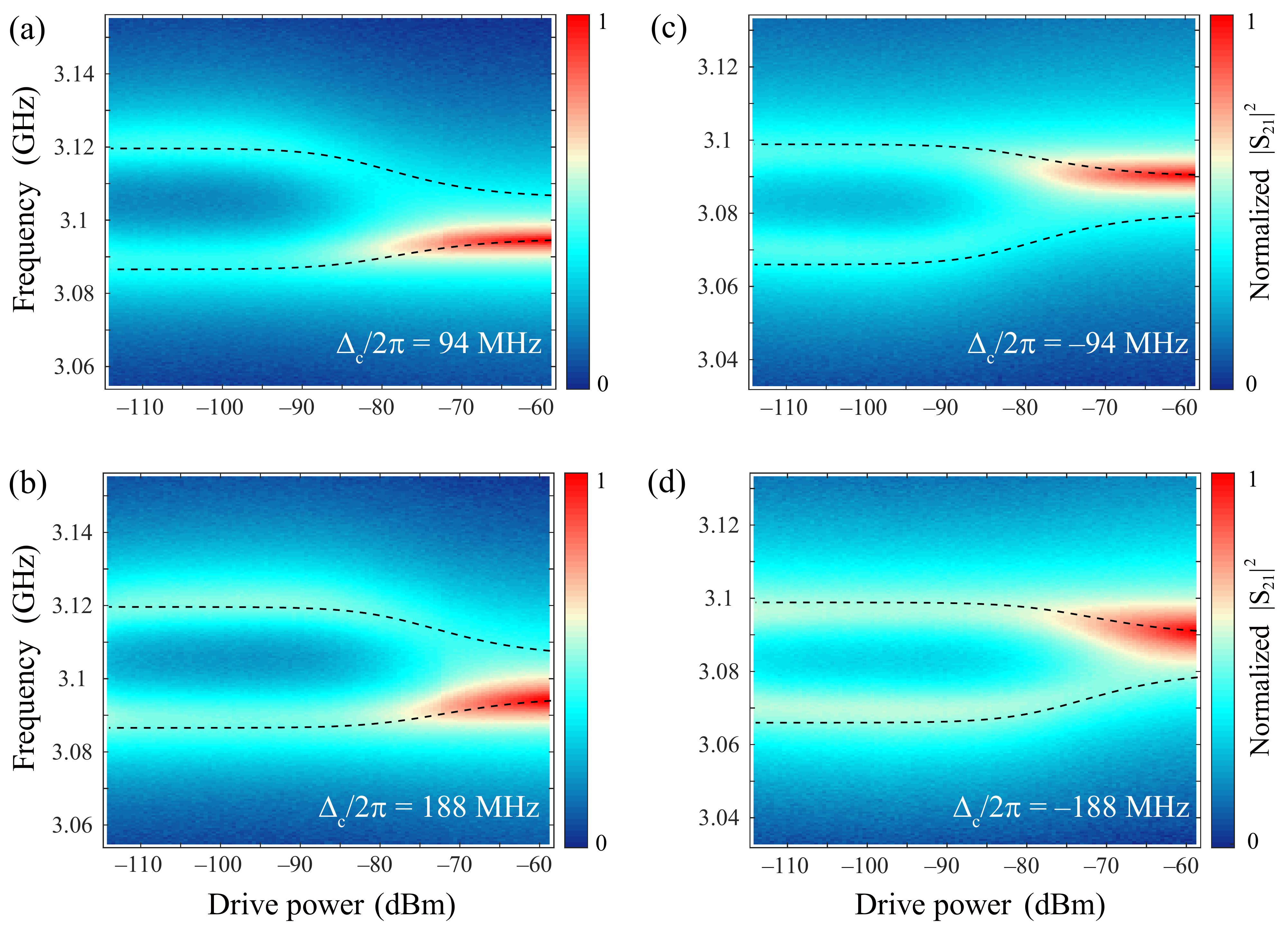}
\caption{The transmission spectra without the EP. (a),(b) The Rabi splittings related to the $s = +$ subensemble, where the drive tone is in resonance with the $s = 0$ and $s=-$ subensembles, respectively. (c),(d) The Rabi splittings related to the $s = -$ subensemble, where the drive tone is in resonance with the $s = 0$ and $s=+$ subensembles, respectively.
The dashed curves are simulated as the real parts of the two eigenvalues, as in Figs.~\ref{fig-cross}(c) and \ref{fig-cross}(d).}
\label{fig-noEP}
\end{figure*}

For the measured transmission spectra in Figs.~\ref{fig-cross}(a) and \ref{fig-cross}(b) (see the main text), the drive tone is tuned to be resonant with the $s=+$ and $-$ subensembles, respectively, while the resonator mode is in resonance with the $s=0$ subensemble (i.e., the resonator mode has a frequency detuning $\Delta_c$ from the drive tone). In the steady state, the drive-power dependence of the reduced magnon occupation $\chi=\langle b^{\dag}b\rangle/N$ as well as the corresponding effective magnon-photon coupling strength $g_{\rm eff,\pm}=g\sqrt{1-2\chi}$ can be described by
\begin{equation}\label{solution2}
\left[(\eta\Delta_{c})^{2}+(\gamma+\eta\kappa)^{2}\right]\chi-\xi\eta(|\Omega_{d}|/\sqrt{N})^2=0,
\end{equation}
where $\eta=g^2(1-2\chi)/(\Delta_{c}^2+\kappa^2)$. The experimental data (circles) in Figs.~\ref{fig-coupling}(a) and \ref{fig-coupling}(b) are the effective magnon-photon couplings extracted from the Rabi splittings in Figs.~\ref{fig-cross}(a) and \ref{fig-cross}(b), respectively. The solid curves are the corresponding effective magnon-photon couplings $g_{\rm eff,+}$ and $g_{\rm eff,-}$ calculated using Eq.~(\ref{solution2}) with $\Delta_{c}/2\pi=-94$ MHz [Fig.~\ref{fig-coupling}(a)] and $\Delta_{c}/2\pi=94$ MHz [Fig.~\ref{fig-coupling}(b)], respectively. It is clear that the numerical results for both $g_{\rm eff,+}$ and $g_{\rm eff,-}$ are in {\it good} agreement with the experimental data of $g_{\rm eff}$. This indicates that the cross relaxation occurs in the P1 centers, which yields $g_{\rm eff,\pm}=g_{\rm eff}$. Therefore, as explained in the main text, $\tilde{\omega}_{c}=\omega_{c}+g^2_{\rm eff,+}/\delta_{+}+g^2_{\rm eff,-}/\delta_{-}$ is reduced to $\omega_c$, because $\delta_{+}=-\delta_{-}$. This gives rise to the {\it symmetric} transmission spectra in Figs.~\ref{fig-cross}(a) and \ref{fig-cross}(b) about the frequency of the resonator mode and the EP becomes observable in the hybrid system.

Similar to Figs.~\ref{fig-cross}(c) and \ref{fig-cross}(d) in the main text, in Fig.~\ref{fig-noEP}, we also show the transmission spectra without the EP, which are measured using an off-resonant drive tone. As in Fig.~\ref{fig-cross}(c), we tune the $s=+$ subensemble to be nearly resonant with the resonator mode. However, instead of driving the $s=+$ subensemble, we use a drive tone to resonantly pump the $s=0$ [Fig.~\ref{fig-noEP}(a)] and $s=-$ [Fig.~\ref{fig-noEP}(b)] subensembles, respectively. Then, as in Fig.~\ref{fig-cross}(d), we tune the $s=-$ subensemble to be nearly resonant with the resonator mode, but use a drive tone to resonantly pump the $s=0$ [Fig.~\ref{fig-noEP}(c)] and $s=+$ [Fig.~\ref{fig-noEP}(d)] subensembles, respectively. The corresponding effective magnon-photon couplings are also calculated using Eq.~(\ref{solution2}) and the two eigenvalues of the considered hybrid system can be then obtained. As in Figs.~\ref{fig-cross}(c) and \ref{fig-cross}(d), the dashed curves are the simulated results for the real parts of the two eigenvalues, which also match the peak positions of the two polariton branches in Fig.~\ref{fig-noEP}.

\section{THE EFFECT OF OTHER SPIN ENSEMBLES ON THE RESONATOR MODE}\label{appendix-E}

When the other two spin ensembles are considered, the effective Hamiltonian of the system becomes
\begin{equation}\label{}
\begin{split}
&H_{\rm eff}=H_{0}+H_{\rm int}+H'_{\rm int},\\
&H_{0}=\omega_{c}a^{\dag}a+\omega_{0}b^{\dagger}b+\omega_{-}c^{\dagger}c+\omega_{+}d^{\dagger}d,\\
&H_{\rm int}=g_{\rm eff}(a^{\dagger}b+ab^{\dagger}),\\
&H'_{\rm int}=g_{\rm eff,-}(a^{\dagger}c+ac^{\dagger})+g_{\rm eff,+}(a^{\dagger}d+ad^{\dagger}),
\end{split}
\end{equation}
where the effective Hamiltonian of the system is written in the standard Hermitian form, without including the decay rates of the resonator mode and the three spin subensembles. In the dispersive regime of $|\omega_{c}-\omega_{\pm}| \gg g_{\rm eff,\pm}$, we can use the Fr\"{o}hlich-Nakajima transformation~\cite{Frohlich55,Nakajima55}, $U=\exp(V)$, with $V^{\dag}=-V$, to eliminate the degrees of freedom of the $s=\pm$ subensembles, where the operator $V$ has the form
\begin{equation}\label{}
V=\lambda_{-}(a^{\dagger}c-ac^{\dagger})+\lambda_{+}(a^{\dagger}d-ad^{\dagger}),
\end{equation}
with
\begin{equation}\label{}
\lambda_{-}=-\frac{g_{\rm eff,-}}{\omega_{c}-\omega_{-}},~~~
\lambda_{+}=-\frac{g_{\rm eff,+}}{\omega_{c}-\omega_{+}}.
\end{equation}

The effective Hamiltonian can be written as
\begin{eqnarray}\label{eff}
\widetilde{H}_{\rm eff}&=&U^{\dag}H_{\rm eff}U  \nonumber\\
           &=&H_{\rm eff}+[H_{\rm eff},V]+\frac{1}{2}[[H_{\rm eff},V],V]+\cdot\cdot\cdot \nonumber\\
    &\approx &H_{0}+(H_{\rm int}+H'_{\rm int}+[H_{0},V])\nonumber\\
     &       &+\Big([H_{\rm int},V]+[H'_{\rm int},V]+\frac{1}{2}[[H_{0},V],V]\Big).
\end{eqnarray}
The first-order terms are
\begin{equation}\label{}
[H_{0},V]=-g_{\rm eff,-}(a^{\dagger}c+ac^{\dagger})-g_{\rm eff,+}(a^{\dagger}d+ad^{\dagger})
         =-H'_{\rm int},\nonumber
\end{equation}
\begin{equation}\label{}
H_{\rm int}+H'_{\rm int}+[H_{0},V]=H_{\rm int}=g_{\rm eff}(a^{\dagger}b+ab^{\dagger}),
\end{equation}
and the second-order terms are
\begin{equation}\label{}
[H_{\rm int},V]=\lambda_{-}g_{\rm eff}(b^{\dagger}c+bc^{\dagger})+\lambda_{+}g_{\rm eff}(b^{\dagger}d+bd^{\dagger}),\nonumber
\end{equation}
\begin{eqnarray}\label{}
[H'_{\rm int},V]&=&2\lambda_{-}g_{\rm eff,-}(c^{\dag}c-a^{\dag}a)
                    +2\lambda_{+}g_{\rm eff,+}(d^{\dag}d-a^{\dag}a)\nonumber\\
                & &+(\lambda_{-}g_{\rm eff,+}+\lambda_{+}g_{\rm eff,-})(c^{\dag}d+cd^{\dag}),
\end{eqnarray}
\begin{eqnarray}\label{}
& &[H_{\rm int},V]+[H'_{\rm int},V]+\frac{1}{2}[[H_{0},V],V]\nonumber\\
&&=[H_{\rm int},V]+\frac{1}{2}[H'_{\rm int},V] \nonumber\\
&&=\lambda_{-}g_{\rm eff,-}(c^{\dag}c-a^{\dag}a)+\lambda_{+}g_{\rm eff,+}(d^{\dag}d-a^{\dag}a) \nonumber\\
&&~~~+\lambda_{-}g_{\rm eff}(b^{\dagger}c+bc^{\dagger})+\lambda_{+}g_{\rm eff}(b^{\dagger}d+bd^{\dagger}) \nonumber\\
&&~~~+\frac{1}{2}(\lambda_{-}g_{\rm eff,+}+\lambda_{+}g_{\rm eff,-})(c^{\dag}d+cd^{\dag}).
\end{eqnarray}
Therefore, the effective Hamiltonian in Eq.~(\ref{eff}) is given by
\begin{eqnarray}\label{}
\widetilde{H}_{\rm eff}&=&H_{\rm eff}^{(0)}+H_{\rm eff}^{(I)}, \nonumber\\
H_{\rm eff}^{(0)}&=&\widetilde{\omega}_{c}a^{\dag}a+\omega_{0}b^{\dagger}b+\widetilde{\omega}_{-}c^{\dagger}c
            +\widetilde{\omega}_{+}d^{\dagger}d+g_{\rm eff}(a^{\dagger}b+ab^{\dagger}), \nonumber\\
H_{\rm eff}^{(I)}
&=&\lambda_{-}g_{\rm eff}(b^{\dagger}c+bc^{\dagger})+\lambda_{+}g_{\rm eff}(b^{\dagger}d+bd^{\dagger})\nonumber\\
&&  +\frac{1}{2}(\lambda_{-}g_{\rm eff,+}+\lambda_{+}g_{\rm eff,-})(c^{\dag}d+cd^{\dag}),
\end{eqnarray}
where
\begin{eqnarray}\label{}
\widetilde{\omega}_{c}&=&\omega_{c}-\lambda_{-}g_{\rm eff,-}-\lambda_{+}g_{\rm eff,+}
                      =\omega_{c}+\frac{g^2_{\rm eff,-}}{\omega_{c}-\omega_{-}}
                                  +\frac{g^2_{\rm eff,+}}{\omega_{c}-\omega_{+}}, \nonumber\\
\widetilde{\omega}_{-}&=&\omega_{-}+\lambda_{-}g_{\rm eff,-}
                       =\omega_{-}-\frac{g^2_{\rm eff,-}}{\omega_{c}-\omega_{-}}, \\
\widetilde{\omega}_{+}&=&\omega_{+}+\lambda_{+}g_{\rm eff,+}
                       =\omega_{+}-\frac{g^2_{\rm eff,+}}{\omega_{c}-\omega_{+}}. \nonumber
\end{eqnarray}
The effective interaction Hamiltonian $H_{\rm eff}^{(I)}$ contains the second-order off-diagonal terms. When neglecting these second-order terms in the dispersive regime $|\omega_{c}-\omega_{\pm}| \gg g_{\rm eff,\pm}$, the effective Hamiltonian of the system becomes
$\widetilde{H}_{\rm eff}\approx H_{\rm eff}^{(0)}=\widetilde{\omega}_{c}a^{\dag}a+\omega_{0}b^{\dagger}b+\widetilde{\omega}_{-}c^{\dagger}c
            +\widetilde{\omega}_{+}d^{\dagger}d+g_{\rm eff}(a^{\dagger}b+ab^{\dagger})$.
Furthermore, we remove the commutative diagonal terms $\widetilde{\omega}_{-}c^{\dagger}c
+\widetilde{\omega}_{+}d^{\dagger}d$ in $H_{\rm eff}^{(0)}$ and include the decay rates of the resonator mode and the $s=0$ subensemble. Then, the reduced Hamiltonian of the hybrid system can be effectively written, in the non-Hermitian form~\cite{Nori19,Nori20}, as
\begin{equation}\label{reduced}
\widetilde{H}_{\rm eff}=(\widetilde{\omega}_{c}-i\kappa)a^{\dag}a+(\omega_{0}-i\gamma)b^{\dagger}b+g_{\rm eff}(a^{\dagger}b+ab^{\dagger}),
\end{equation}
with
\begin{equation}
\widetilde{\omega}_{c}=\omega_{c}+\frac{g^2_{\rm eff,-}}{\omega_{c}-\omega_{-}}
                                  +\frac{g^2_{\rm eff,+}}{\omega_{c}-\omega_{+}}.
\end{equation}
The Hamiltonian (\ref{reduced}) has the same form as Eq.~(\ref{matrix}) in the main text and its two eigenvalues are also given by  Eq.~(\ref{eigenvalues}) in the main text, i.e., $\omega_{1,\,2}$ in Eq.~(\ref{eigen}), but $\omega_{c}$ therein is replaced by $\widetilde{\omega}_{c}$.

\end{document}